\documentclass{emulateapj}

\usepackage[]{natbib}
\usepackage[]{graphicx}
\usepackage[]{subfigure}
\usepackage{color}

\usepackage{txfonts}

\shorttitle{Temporal properties of GX~301-2}
\shortauthors{Evangelista et al.}

\begin{document}


\title{Temporal properties of GX~301-2 over a year-long observation with SuperAGILE}

  \author{{Y.~Evangelista}\altaffilmark{1,2,*}, 
    M.~Feroci\altaffilmark{1},  
    E.~Costa\altaffilmark{1}, 
    E.~Del~Monte\altaffilmark{1}, 
    I.~Donnarumma\altaffilmark{1},
    I.~Lapshov\altaffilmark{1}, 
    F.~Lazzarotto\altaffilmark{1}, 
    L.~Pacciani\altaffilmark{1}, 
    M.~Rapisarda\altaffilmark{3}, 
    P.~Soffitta\altaffilmark{1},
    A.~Argan\altaffilmark{1}, 
    G.~Barbiellini\altaffilmark{4,5}, 
    F.~Boffelli\altaffilmark{6}, 
    A.~Bulgarelli\altaffilmark{7}, 
    P.~Caraveo\altaffilmark{8},
    P.W.~Cattaneo\altaffilmark{6}, 
    A.~Chen\altaffilmark{8}, 
    F.~D'Ammando\altaffilmark{1,9}, 
    G.~Di~Cocco\altaffilmark{7}, 
    F.~Fuschino\altaffilmark{7}, 
    M.~Galli\altaffilmark{10},
    F.~Gianotti\altaffilmark{7}, 
    A.~Giuliani\altaffilmark{8}, 
    C.~Labanti\altaffilmark{7}, 
    P.~Lipari\altaffilmark{2},
    F.~Longo\altaffilmark{4,5}, 
    M.~Marisaldi\altaffilmark{7}, 
    S.~Mereghetti\altaffilmark{8}, 
    E.~Moretti\altaffilmark{4,5}, 
    A.~Morselli\altaffilmark{11}, 
    A.~Pellizzoni\altaffilmark{12},
    F.~Perotti\altaffilmark{8}, 
    G.~Piano\altaffilmark{1,9}, 
    P.~Picozza\altaffilmark{9, 11}, 
    M.~Pilia\altaffilmark{12,13}, 
    M.~Prest\altaffilmark{13}, 
    G.~Pucella\altaffilmark{3}, 
    A.~Rappoldi\altaffilmark{6},
    S.~Sabatini\altaffilmark{1,11}, 
    E.~Striani\altaffilmark{9,11}, 
    M.~Tavani\altaffilmark{1,9}, 
    M.~Trifoglio\altaffilmark{7}, 
    A.~Trois\altaffilmark{1}, 
    E.~Vallazza\altaffilmark{4},
    S.~Vercellone\altaffilmark{14}, 
    V.~Vittorini\altaffilmark{1}, 
    A.~Zambra\altaffilmark{15},
    L.~A.~Antonelli\altaffilmark{16,17}, 
    S.~Cutini\altaffilmark{16,18}, 
    C.~Pittori\altaffilmark{16,18}, 
    B.~Preger\altaffilmark{16,18}, 
    P.~Santolamazza\altaffilmark{16,18},
    F. Verrecchia\altaffilmark{16,18}, 
    P.~Giommi\altaffilmark{16, 19}, 
    L. Salotti\altaffilmark{19}}

\altaffiltext{1}{INAF IASF Roma, Via Fosso del Cavaliere 100, I-00133 Roma, Italy}
\altaffiltext{2}{Dip. di Fisica, Universit\`a degli Studi di Roma ``La Sapienza'', P.le A. Moro 5, I-00185 Roma, Italy}
\altaffiltext{3}{ENEA, Via E. Fermi 45, I-00044 Frascati (Rm), Italy}
\altaffiltext{4}{INFN Trieste, Padriciano 99, I-34012 Trieste, Italy}
\altaffiltext{5}{Dip. di Fisica, Universit\`a di Trieste, Via Valerio 2, I-34127 Trieste, Italy}
\altaffiltext{6}{INFN Pavia, Via Bassi, 6 I-27100 Pavia, Italy}
\altaffiltext{7}{INAF IASF Bologna, Via Gobetti 101, I-40129 Bologna, Italy}
\altaffiltext{8}{INAF IASF Milano, Via E. Bassini 15, I-20133 Milano, Italy}
\altaffiltext{9}{Dip. di Fisica, Universit\`a degli Studi di Roma ``Tor Vergata'',  Via della Ricerca Scientifica 1, I-00133 Roma, Italy}
\altaffiltext{10}{ENEA C.R. ``E. Clementel'', Via don Fiammelli 2, I-40128 Bologna, Italy}
\altaffiltext{11}{INFN Roma Tor Vergata, Via della Ricerca Scientifica 1, I-00133, Roma, Italy}
\altaffiltext{12}{INAF Osservatorio Astronomico di Cagliari, loc. Poggio dei Pini, strada 54, I-09012, Capoterra (Ca), Italy}
\altaffiltext{13}{Dip. di Fisica e Matematica, Universit\`a dell'Insubria, Via Valleggio 11, I-20100 Como, Italy}
\altaffiltext{14}{INAF IASF Palermo, Via U.\ La Malfa 153, I-90146 Palermo, Italy}
\altaffiltext{15}{Consorzio Interuniversitario per la Fisica Spaziale, Viale Settimio Severo 63, I-10133 Torino, Italy}
\altaffiltext{16}{ASI Science Data Center, Via G.\ Galilei, I-00044 Frascati (Rm), Italy}
\altaffiltext{17}{INAF Osservatorio Astronomico di Roma, Via di Frascati 33, I-00040 Monte Porzio Catone (Rm), Italy}
\altaffiltext{18}{INAF personnel resident at ASI Science Data Center}
\altaffiltext{19}{Agenzia Spaziale Italiana, Unit\`a Osservazione dell'Universo, Viale Liegi 26, 00198 Roma, Italy}
\altaffiltext{*}{yuri.evangelista@iasf-roma.inaf.it}



\begin{abstract}
  We present the long-term monitoring of the High Mass X-ray
  Binary GX~301-2 performed  with the SuperAGILE instrument
  on-board the {\it AGILE} mission.
  The source was monitored in the 20--60~keV energy band during the first year of the 
  mission from 2007 July 17 to 2008 August 31, covering about one
  whole orbital period and three more pre-periastron passages for a
  total net observation time of about 3.7~Ms. The SuperAGILE dataset
  represents one of the most continuous and complete monitoring at
  hard X-ray energies of the 41.5~day long binary period available to
  date.

  The source behavior was characterized at all orbital phases in
  terms of hard X-ray flux, spectral hardness, spin period history,
  pulsed fraction and pulse shape profile.  We also complemented the
  SuperAGILE observations with the soft X-ray data of the
  RossiXTE/ASM. Our analysis shows a clear orbital modulation of the
  spectral hardness, with peaks in correspondence with the
  pre-periastron flare and near phase 0.25. The hardness peaks we
  found could be related with the wind-plus-stream accretion model
  proposed in order to explain the orbital light
  curve modulation of GX~301-2.

  Timing analysis of the pulsar spin period shows that the secular
  trend of the $\sim$680~s pulse period is consistent with the
  previous observations, although there is evidence of a slight
  decrease in the spin-down rate.
  The analysis of the hard X-ray pulsed emission also showed a
  variable  pulse shape profile as a function of the orbital phase,
  with substructures detected near the passage at the periastron, and
  a clear modulation of the pulsed fraction, which appears in turn
  strongly anti-correlated with the source intensity.

\end{abstract}


\section{Introduction}

GX~301-2 is a High Mass X-ray Binary (HMXB) system containing an
X-ray pulsar, with a $\sim$685~s rotation period, orbiting the
B-emission line hypergiant star (B1 Ia+) Wray 977. The X-ray source
was discovered in 1969 April \citep{Lewin1971} with a balloon
observation, subsequently \cite{White1976} discovered the 700~s
X-ray pulsation. The pulsar orbit has an eccentricity of  0.462
\citep{Koh1997} and a period of $\sim$41.5~days  
\citep{White1978, Sato1986, Koh1997, Doroshenko2008}.
The mass function of the system is 31.1~M$_\odot$ with a companion
mass in the range 39~$<$~M~$<$~53~M$_\odot$, estimated by the measurement
of optical radial velocity amplitude \citep{Kaper2006}.
The radius of Wray 977 obtained by  \cite{Kaper2006} by fitting
atmosphere model is 62~$R_\odot$, while the effective temperature is
about 1.8~$\times$~10$^4$~${\rm K}$.

GX~301-2 shows regular X-ray flares about two days before the periastron
passage, while a secondary intensity peak was sometimes observed
near the apastron passage \citep{Pravdo1995, Koh1997, Pravdo2001}.
Assuming a distance of 3~kpc, as derived by \cite{Kaper2006},
the source luminosity during pre-periastron (PP) flares reaches
values of $\sim 10^{37}$~erg$ \: $s$^{-1}$, about 25 times larger
than the luminosity showed by the source in the other orbital
phases.

Since its discovery, GX~301-2 exhibited unpredictable torque
variations superimposed on a secular spin trend, which passed from
a spin-up to a spin-down state between MJD~48500 and MJD~49500.
Two rapid spin-up episodes with $\dot{\nu} \sim 2\mbox{--}5 \times
10^{-12}\, {\rm Hz}\, {\rm s}^{-1}$ were also found by \cite{Koh1997}
analyzing BATSE data, thus confirming the erratic behavior of the
pulsar spin period. \cite{Koh1997} suggested that the long-term
spin-up trend observed between 1984 and 1994 might not be regular and
continuous, but due to several rapid spin-up episodes, which in turn
could be related to the formation of a temporary accretion
disk. \cite{Pravdo2001}, using BATSE data, found evidence that the
hard  X-ray (20-50~keV) intensity of the binary system during
quiescent orbital phases is correlated with the pulse period, suggesting
a possible relation between the spin period behavior and the mass
accretion rate. 

Several wind accretion models were proposed to explain both the
regular PP and apastron flaring activities of the source. A dense
disk around Wray~977, which intercepts the neutron star (NS) orbit
near the periastron and the apastron, was proposed by
\cite{Pravdo1995} and \cite{Pravdo2001}. However, this model seems
unable to provide a satisfactory modeling of the observed orbital
X-ray light curve of the source \citep{Leahy2002} and optical
observations by \cite{Kaper2006} found no confirmation of the
presence of a disc. The existence of a high density stream of
matter in addition to the stellar wind was proposed by
\cite{Haberl1991, Leahy1991, Leahy2002,  Leahy2008}. 
The stream outflows from Wray~977 due to tidal interaction 
intercepts the NS orbit near the periastron and the apastron,
causing a large increase in the mass accretion rate with a sudden
enhancement of the X-ray luminosity from the neutron star.
Signatures of the presence of such a stream were found by
\cite{Saraswat1996} analyzing the low-energy excess in the {\it ASCA}
spectrum of GX~301-2, and also by \cite{Kaper2006} from the
orbital modulation of the spectral lines detected in the stellar
wind of the optical companion. The wind-plus-stream model predicts
also a large variation in the column density as a function of the
orbital phase. The $N_{\mathrm{H}}$ is expected to change from a value of
$\sim 10^{23}\,$cm$^{-2}$ at orbital phase 0.8 to values greater
than 10$^{24}\,$cm$^{-2}$ during the PP flares and between orbital
phases 0.2--0.3.

We observed the HMXB GX~301-2 with the SuperAGILE (SA) experiment
onboard the  Astro-rivelatore Gamma ad Immagini LEggero ({\it AGILE}) satellite during the first year of the mission. 
{\it AGILE}, launched on
2007 April 23, is a scientific mission of the Italian Space
Agency dedicated to high-energy astrophysics \citep{Tavani2009}.
In the following sections we report on the analysis of the
SuperAGILE data, in the energy range 20--60 keV. In
Section~\ref{sec:obs_and_data} we describe the SuperAGILE experiment
and the observations of GX~301-2 we performed, in
Section~\ref{sec:timing} we present the timing analysis of the pulsar
emission and in Section~\ref{sec:orb_lc} we analyze the orbital light
curve of GX~301-2 using the hard X-ray SuperAGILE and the soft X-ray
{\it RXTE}/ASM data and the orbital variability of the source pulsed
fraction. Discussion of the data analysis and the interpretation of
the results are presented in Section~\ref{sec:discussion}, while in
Section~\ref{sec:conclusion} we draw our conclusions.

\section{SuperAGILE observations and data reduction}
\label{sec:obs_and_data}
\subsection{The SuperAGILE Experiment}
SuperAGILE (see \citealt{Feroci2007} for an extensive description of
the experiment) is the hard X-ray monitor of the {\it AGILE} mission and
is based on four one-dimensional coded-mask instruments, each one
equipped with a $\sim360\,$cm$^2$ silicon microstrip detector and a
tungsten mask-collimator system. 
The experiment has $\sim$1~sr field of view (FoV), 6~arcmin
angular resolution, and a point source location accuracy better
than 2~arcmin for intense sources. SuperAGILE is able to detect
X-ray photons between 18~keV and 60~keV with a resolution of
$\sim$8~keV FWHM and an on-axis effective area of 250~cm$^2$. The
time resolution of the instrument is 2~$\mu$s while the accuracy
is about 5~$\mu$s. The dead time is negligible in normal operation
conditions, except for the passages through the South Atlantic Geomagnetic
Anomaly where the vetoes from the {\it AGILE} anticoincidence systems
reduce the SA live time to $\sim$10\% or less. In the standard
analysis this fraction of the orbit is usually removed from the
data.

In standard operation mode SuperAGILE provides event by event
data, from which position and normalized count rates for each
source detected in the FoV are derived. The data are downloaded
from the spacecraft about every 100 minutes during the AGILE
passage over the Malindi ground station, sent to the ASI Science
Data Center (ASDC) in Rome and then processed by the SuperAGILE
Scientific Pipeline (SASOA) developed by the SuperAGILE team at
INAF/IASF Rome \citep{Lazzarotto2008} and running both at the ASDC
and at IASF. The automated analysis of the SuperAGILE data
autonomously updates a source catalog and orbital light curves for
bright sources are then posted at the ASDC web
site\footnote{http://agile.asdc.asi.it/sagilecat\_sources.html}.

\subsection{SuperAGILE Observations of GX~301-2}

SuperAGILE observations of the HMXB GX~301-2 started in mid 2007 July,
just after the end of the {\it AGILE} Commissioning Phase, as part 
of the Science Verification Phase. Indeed, GX~301-2 was the first
light of the experiment. Due to the solar constraints and  the
{\it AGILE} pointing
strategy\footnote{http://agile.asdc.asi.it/current\_pointing.html}
the satellite is operated by performing long observations,
typically of 2--4 weeks duration, during which the pointing
direction slowly drifts (at a rate of $\sim$1$^{\circ}$~day$^{-1}$).
SuperAGILE performed six long observations of GX~301-2,
for a total net exposure time of about 3.7~$\times$~10$^6$~s.
Table~\ref{tab:pointings} shows the journal of the observations.
The first observation and part of the
second one were performed during the {\it AGILE} Science Verification
Phase that started in mid 2007 July and ended on 2007 November
30. The other observations were carried out during the
AGILE AO1 program, which started on 2007 December 1 and
lasted one year. The last column in Table~\ref{tab:pointings} 
shows the source off-axis angle in the SuperAGILE FoV for each
observation. The configuration of SuperAGILE implies that
the source counts collected by the instrument strongly depend on
the off-axis angle \citep[see e.g.][]{Feroci2007}, and thus also the
detection significance is dependent on the source position in the
FoV.
\begin{figure*}[!t]
 \centering
  \subfigure{
     \includegraphics[width=0.23\textwidth]{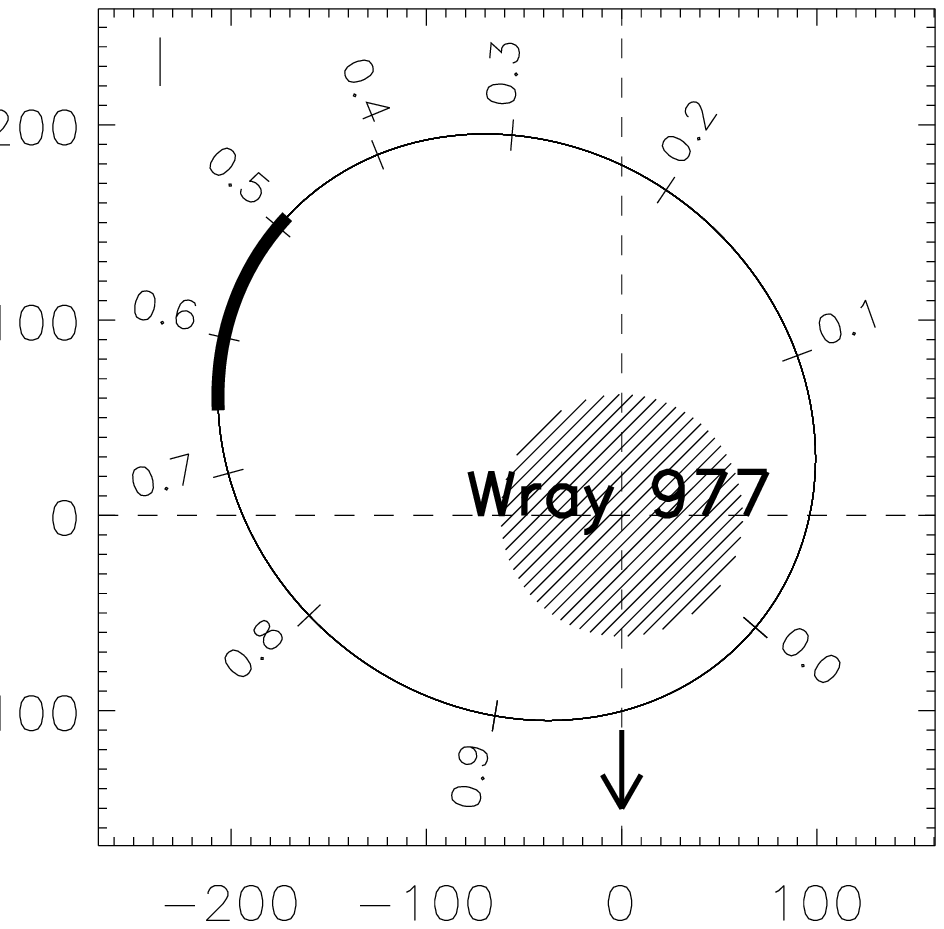}
 }
  \subfigure{
     \includegraphics[width=0.23\textwidth]{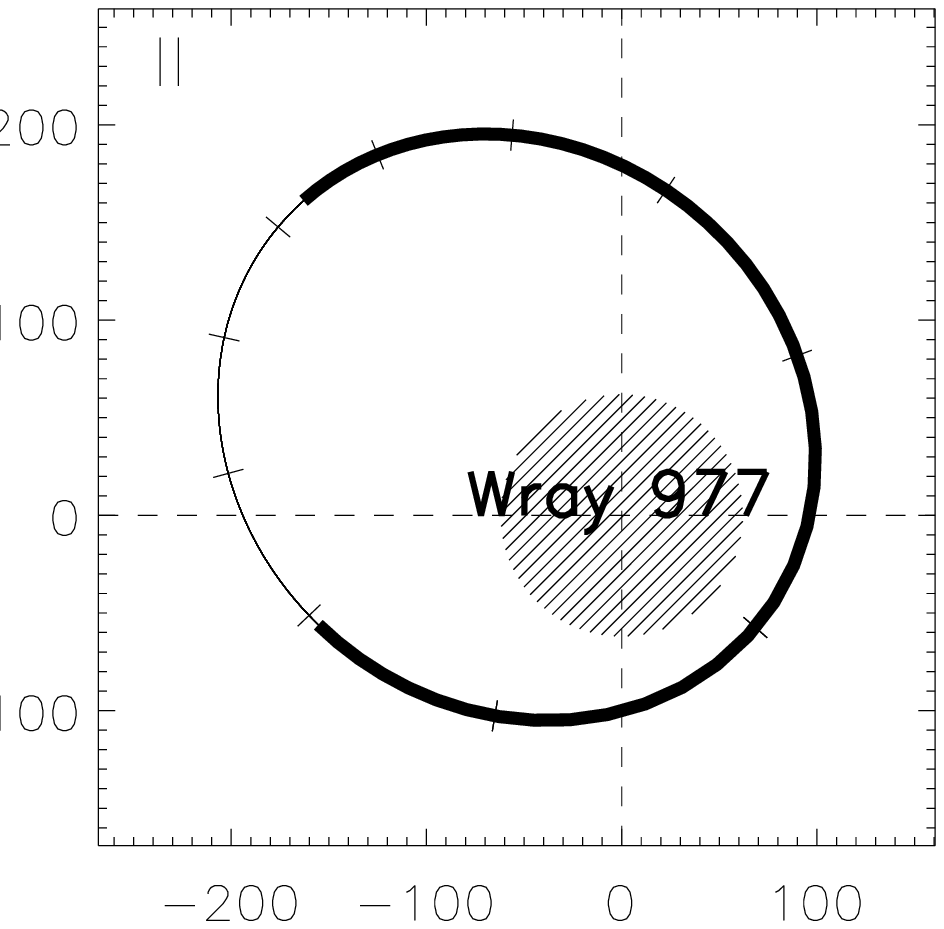}
 }
  \subfigure{
     \includegraphics[width=0.23\textwidth]{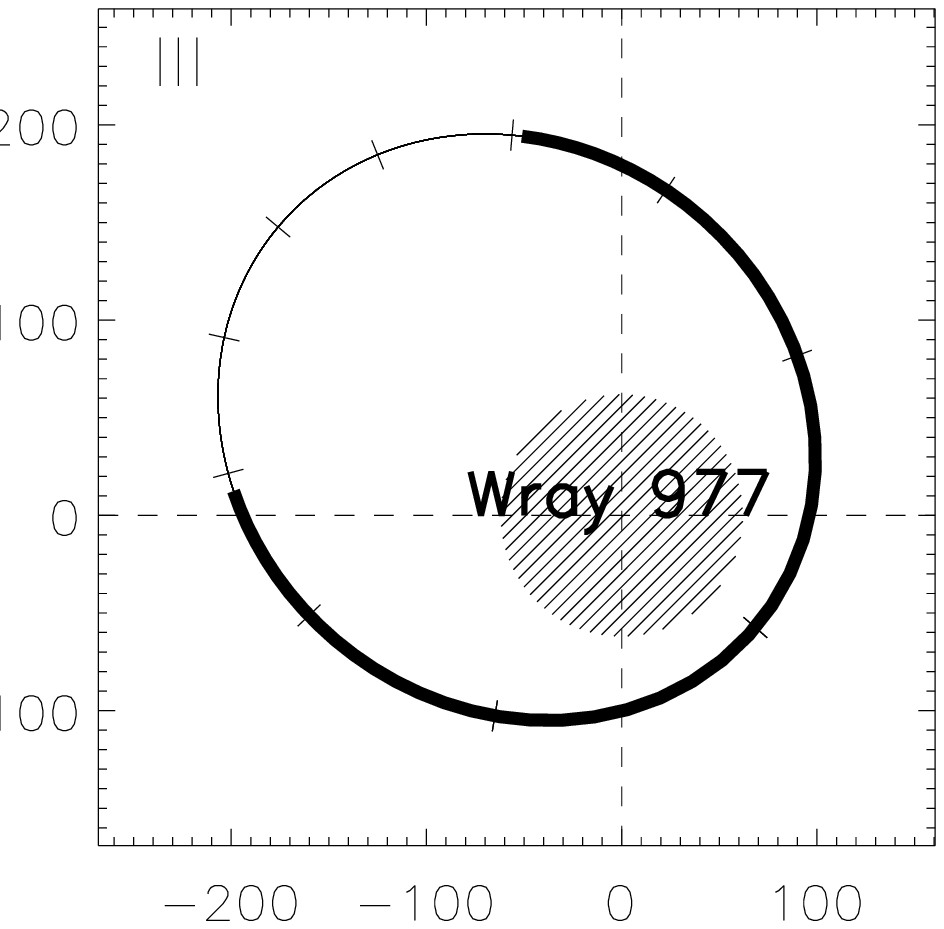}
 }
  \subfigure{
     \includegraphics[width=0.23\textwidth]{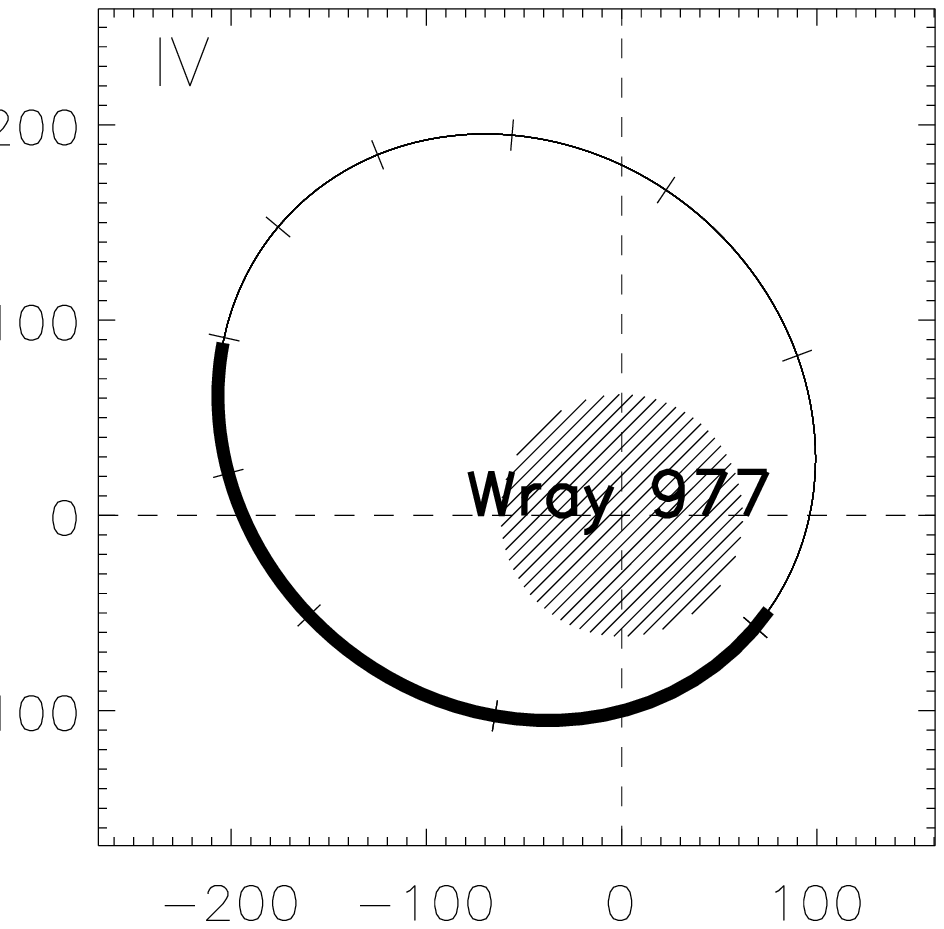}
 }
  \subfigure{
     \includegraphics[width=0.23\textwidth]{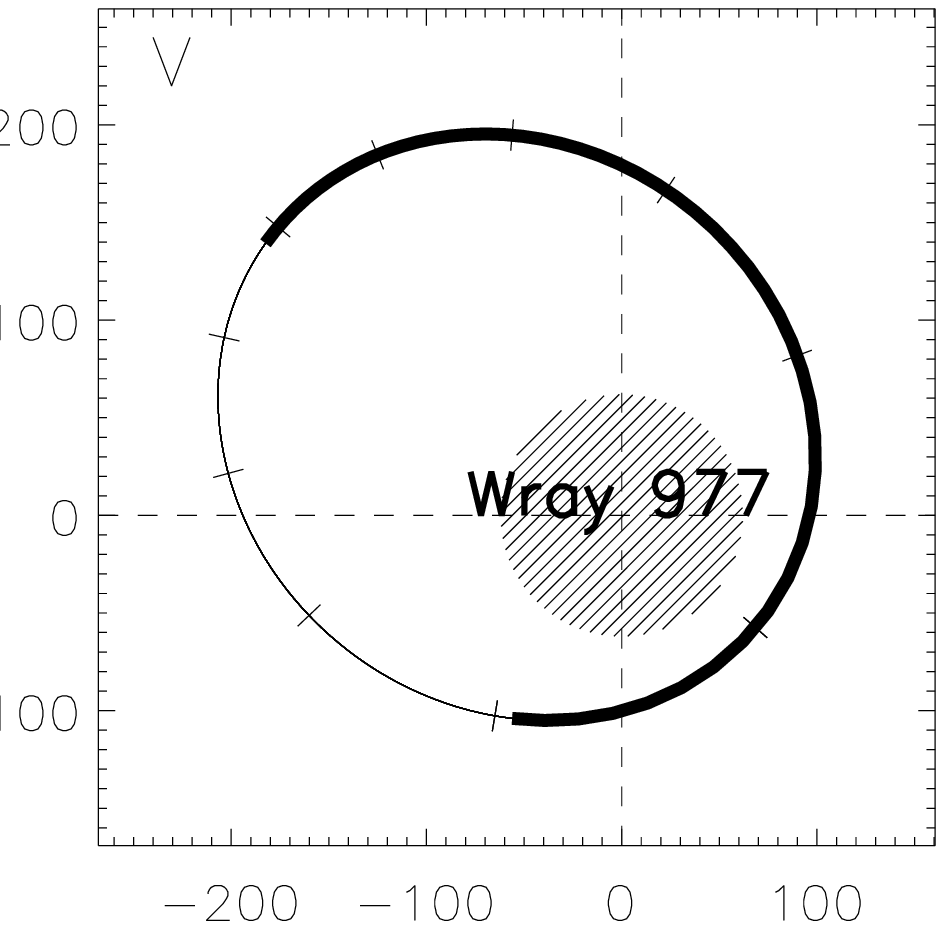}
 }
  \subfigure{
     \includegraphics[width=0.23\textwidth]{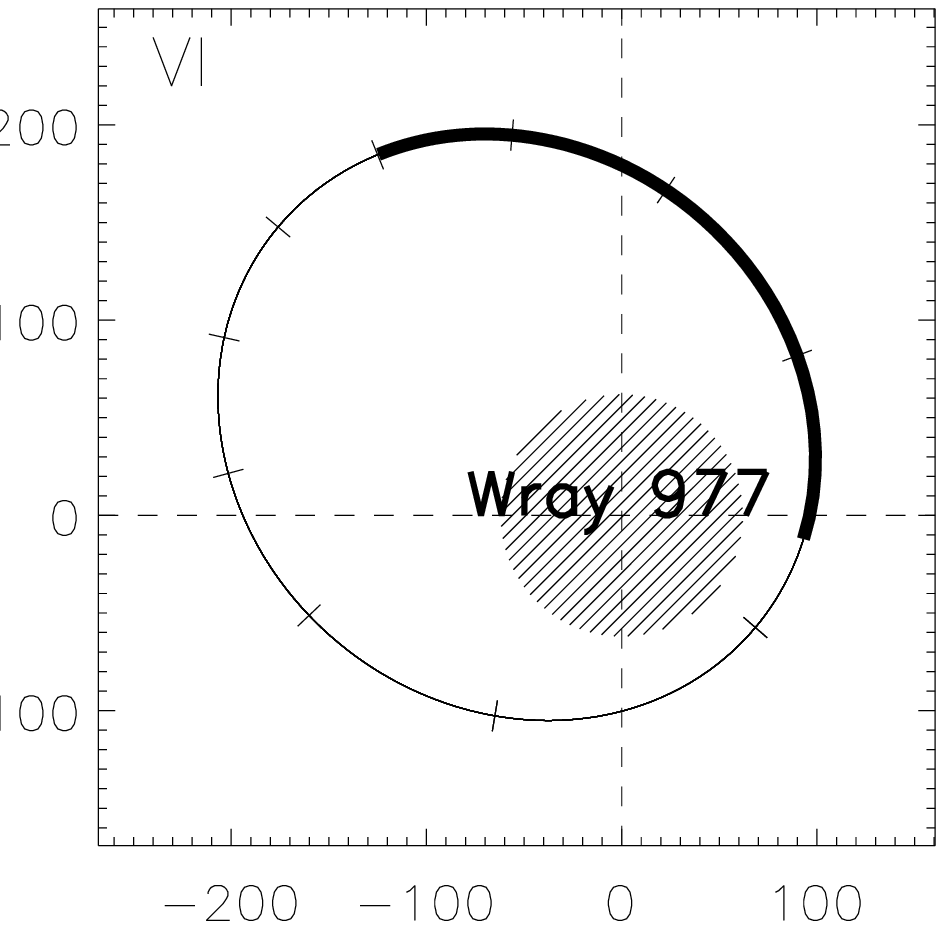}
 }
   \caption{\label{fig:ph_observed}Graphical representations of the GX~301-2 binary
     system. The solid thick line in each panel shows the orbital
     phases observed by SuperAGILE during the six observation blocks
     summarized in Table~\ref{tab:pointings}. Ticks on the Neutron
     Star orbits represent the orbital phases with step 0.1 as
     displayed in the upper-left panel, the NS orbital motion is
     counter-clockwise. Units along the axis are in solar radii while
     the arrow in the upper-left panel represents the observer
     direction. Roman numerals in each panel indicate the Observation
     ID.}

\end{figure*}

The SuperAGILE observations took place in different segments of the
GX~301-2 binary orbit. 
In Figure~\ref{fig:ph_observed} we graphically show the orbital
phases of our six observations: three of them covered a large fraction
of a complete binary orbit (Observations ID II, III and V), while
the other three were shorter, and covered the equivalent of one full
binary orbit all together, but arranged over three different orbits.
The SuperAGILE data thus allow for a often continuous monitoring of the
complete 41.5 days binary cycle. It is worth noting that the monitoring of
four different passages at the pre-periastron orbital phase (0.95) was
included in the SuperAGILE observations. 

\begin{table*}
  \begin{center}
  \caption{\label{tab:pointings} Journal of the {\it AGILE} Observations of GX~301-2. } 
  \begin{tabular}{cccccc}
    \small{Obs. ID} & \small{Mission Phase} &\small{MJD Start} &
    \small{MJD Stop} & \small{Orbital Phase} & \small{Off-axis Angle (deg)}\\
    \tableline\tableline
    I   & SVP        & 54298.4 & 54305.3 & 0.49--0.66 & 7--10\\
    II  & SVP, AO1   & 54311.6 & 54339.1 & 0.81--0.47 & 4--15\\
    III & AO1        & 54473.6 & 54497.7 & 0.71--0.29 & 5--18\\
    IV  & AO1        & 54510.6 & 54527.3 & 0.60--0.01 & 10--17\\
    V   & AO1        & 54647.6 & 54672.9 & 0.91--0.52 & 14--20\\
    VI  & AO1        & 54694.3 & 54709.5 & 0.03--0.40 & 12--17\\
    \tableline
  \end{tabular}
  \tablecomments{First column indicates the Observation ID, the second column specifies the {\it AGILE} mission phase:
    SVP=Science Verification Phase, AO1=Scientific Observations
    Program pointings (see {\it AGILE} Mission Announcement of Opportunity
    Cycle-1: http://agile.asdc.asi.it/).}
  \end{center}
\end{table*}

\subsection{Data Reduction and Analysis}

\begin{figure*}[!t]
 \centering
  \includegraphics[width=0.99\textwidth] {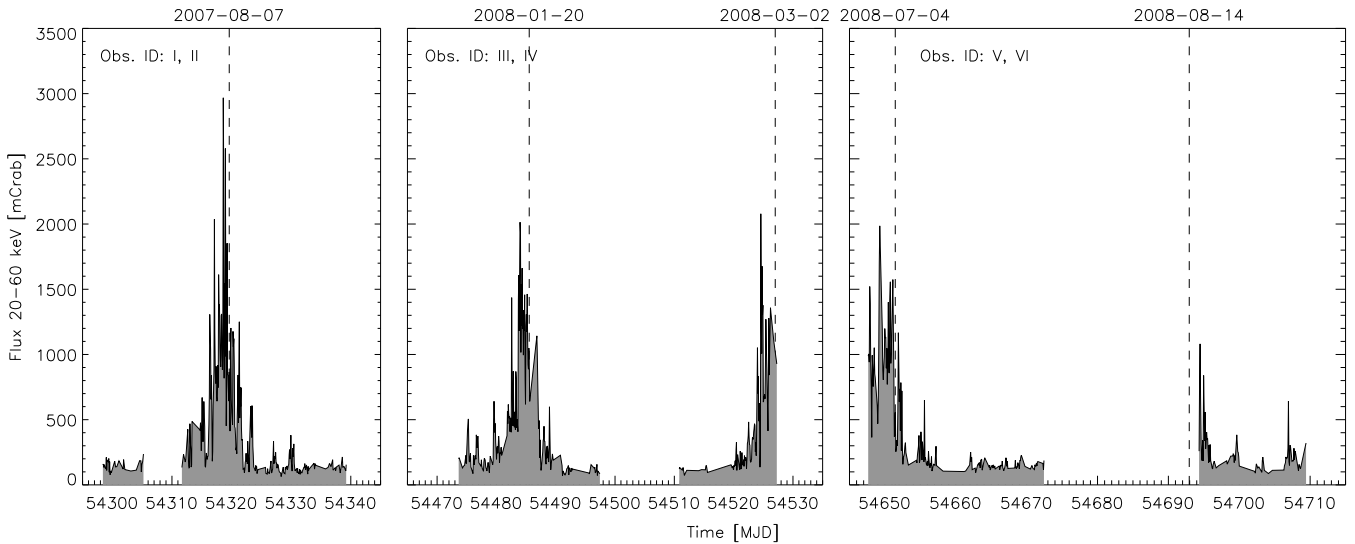}
   \caption{\label{fig:total_data}Light curve of GX~301-2 as observed by SuperAGILE. The
     source was monitored for MJD~54298 to MJD~54709. Dashed lines
     indicate the periastron passages on the basis of the orbital
     parameters calculated by \cite{Doroshenko2008}.
   }
\end{figure*}

SuperAGILE data reduction was performed using the standard SASOA
pipeline \citep{Lazzarotto2008}, which extracts high level
products from the photon-by-photon SuperAGILE data. Production of
source images was performed with the SuperAGILE Enhanced Multi
Imaging (EMI) procedure \citep{Evangelista2008} which allows fast
and accurate analysis for bright sources. For each {\it AGILE} orbit the
source count rate was automatically extracted in the energy range
20--60~keV, excluding the time spent by the satellite in crossing
the South Atlantic Geomagnetic Anomaly and the time when the Earth
occulted the line of sight to the source. Normalized count rates
(in units of counts~cm$^{-2}\,$s$^{-1}$) were then obtained by
considering the effective area of the exposed portion of the
detectors and assuming a Crab-like spectrum
\citep[e.g.][]{Frontera2007}. Fluxes in unit of the Crab flux were  
calculated by applying a normalization factor of 0.15 to the
normalized count rates. Uncertainties on the flux measurements
were estimated by adding to the statistical error a systematic
error equal to the 8\% of the measured flux \citep{Feroci2009}. In order to avoid 
the introduction of biases in the light curve we did not select
the source detections for significance but all the observations
were used in the analysis. A detailed description of the method
used for the generation of the SuperAGILE light curves can be
found in \cite{Feroci2009}. The resulting 20--60~keV light curve of 
GX~301-2 with a $\sim$~6~$\times$~10$^3$~s binning is shown in
Figure~\ref{fig:total_data}.

SuperAGILE events are time-tagged with microsecond precision by
the {\it AGILE} on-board Pulse-Per-Second (PPS) system, which is
synchronized to UTC once a second by the Global Positioning System
(GPS). In order to perform timing analysis all the SuperAGILE
events are  converted on ground to the Terrestrial Dynamical Time
(TDT) reference system. The photon arrival time (TOA) is then
converted to Barycenter Dynamical Time and the correction for the
arrival delay at the Solar System Barycenter is 
calculated by using the JPL planetary ephemeris DE405. A more detailed
description of the correction algorithm may be found in
\cite{Pellizzoni2009}. 
Correction for the time delay introduced by the binary orbital
motion was performed using the Blandford-Teukolsky model
\citep{Bladford1976, Taylor1989}. To perform the TOA
correction we used the orbital parameters provided by
\cite{Koh1997}, except for the orbital period ($P_{\mathrm{orb}}$)
and the time of periastron passage ($T_0$), for which we utilized
the updated values derived by \cite{Doroshenko2008}. 
These parameters are given in Table~\ref{tab:orbital_parameters}.
The quoted uncertainties are given at 1~$\sigma$ level throughout the
paper, unless otherwise specified.

\begin{table}[h]
  \begin{center}
    \caption{\label{tab:orbital_parameters} Orbital Parameters of GX~301-2.}
    \begin{tabular}{lccc}
      Parameter & \cite{Koh1997}  &    \cite{Doroshenko2008}     &  \\
      \tableline\tableline
      $P_{\mathrm{orb}}$ (day)      & 41.498   $\pm$ 0.006 &  41.482 $\pm$ 0.006  \\
      $e$                           & 0.462    $\pm$ 0.042 &    ...                \\
      a$_x $sin$\,i$   (lt-s)     & 368.3    $\pm$ 11.1  &    ...                 \\
      $\omega$         (deg)      & 310.4    $\pm$ 4.2   &    ...                 \\
      T$_{0}$           (MJD)     & 48802.79 $\pm$ 0.36  & 53531.63 $\pm$ 0.03 \\
      \tableline
    \end{tabular}
    \tablecomments{ Parameters are derived by~\cite{Koh1997} and by
      \cite{Doroshenko2008}. The uncertainties are at the 3$\sigma$ level
      both for the Koh and the Doroshenko parameters.} 
  \end{center}
\end{table}

\section{Analysis of the spin period}
\label{sec:timing}

\subsection{Pulse-shape Profile}

A search for the pulsed emission from the NS was performed on the
corrected events of all the SuperAGILE source observations using
the standard epoch folding technique \citep{Leahy1983}. Due to the
different off-axis angles to which the source was observed in the
different pointings (see Table \ref{tab:pointings}), corresponding
to significantly different exposed effective areas, for each orbital
phase bin we selected the available data with the highest
signal-to-noise ratio, in order to obtain the best determination
of the pulse shape. In case of pulse shape variability between
different orbital cycles, this choice biases the pulse shape to
the specific selection. However, we verified that in the other
SuperAGILE observations of the same orbital phase bin the pulse shape
was consistent, within the statistical uncertainties. Following this
approach, pulse shape profiles for orbital phases between 0.26 and
0.64 and between 0.81 and 0.06 were obtained from data collected
during Observation I and II, while the other pulse shape profiles
were built using data taken during Observation III.
The width of the phase intervals was chosen according to the
signal-to-noise ratio in the profiles. This implies to have a
phase bin equal to $\sim$0.10 ($\sim$4~days) for the orbital
phase interval 0.06--0.34, where the source flux is lower,
while a phase bin size $\lesssim$0.05 was used for the other
orbital phases with higher signal-to-noise.
We were not able to detect NS pulsation in orbital intervals 
0.16--0.26, 0.34--0.38, 0.43--0.49, 0.54--0.59 and
0.64--0.71.

\begin{figure*}[!t]
 \centering
 \subfigure{
     \includegraphics[width=0.30\textwidth]{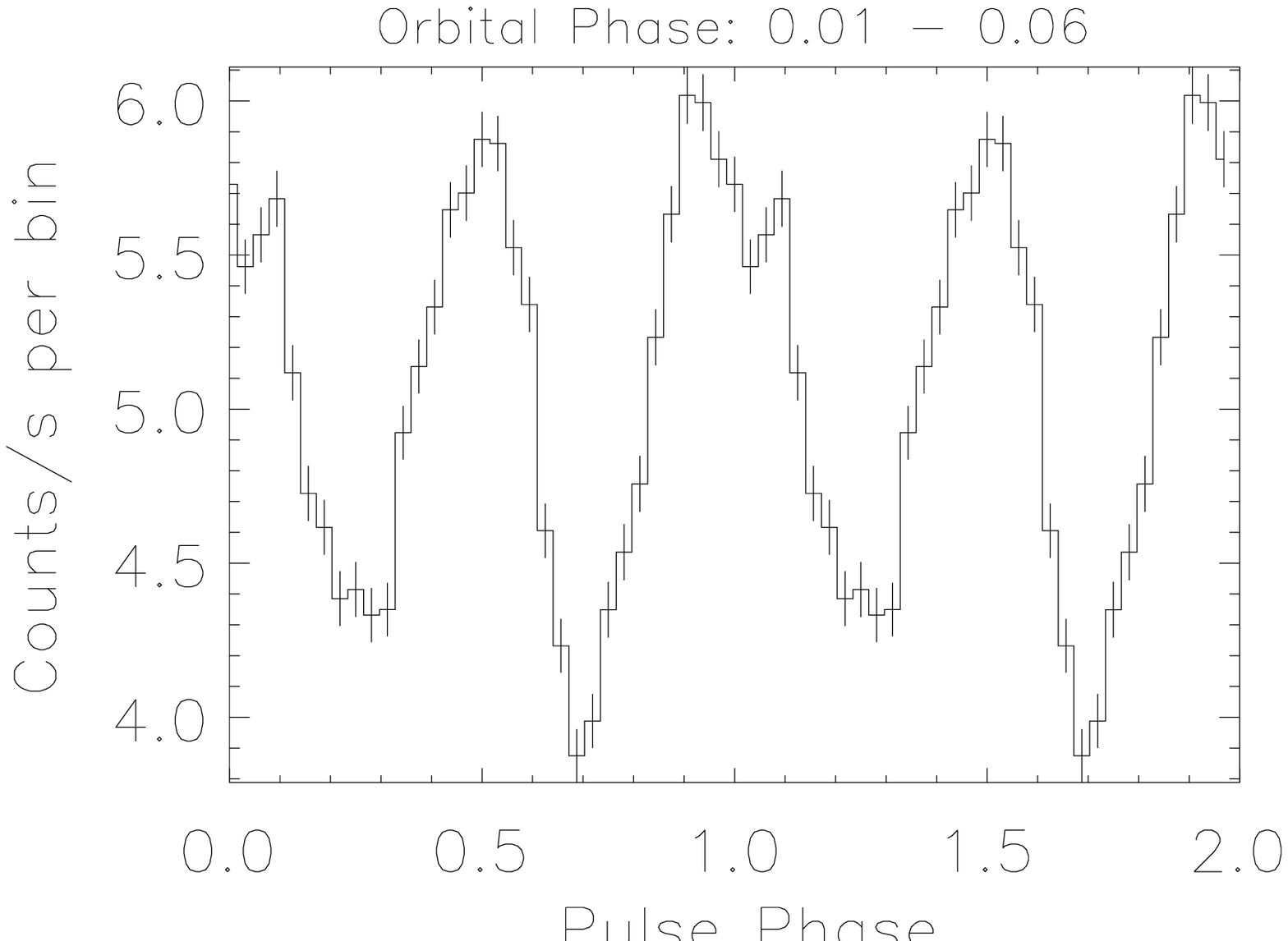}
 }
 \subfigure{
     \includegraphics[width=0.30\textwidth]{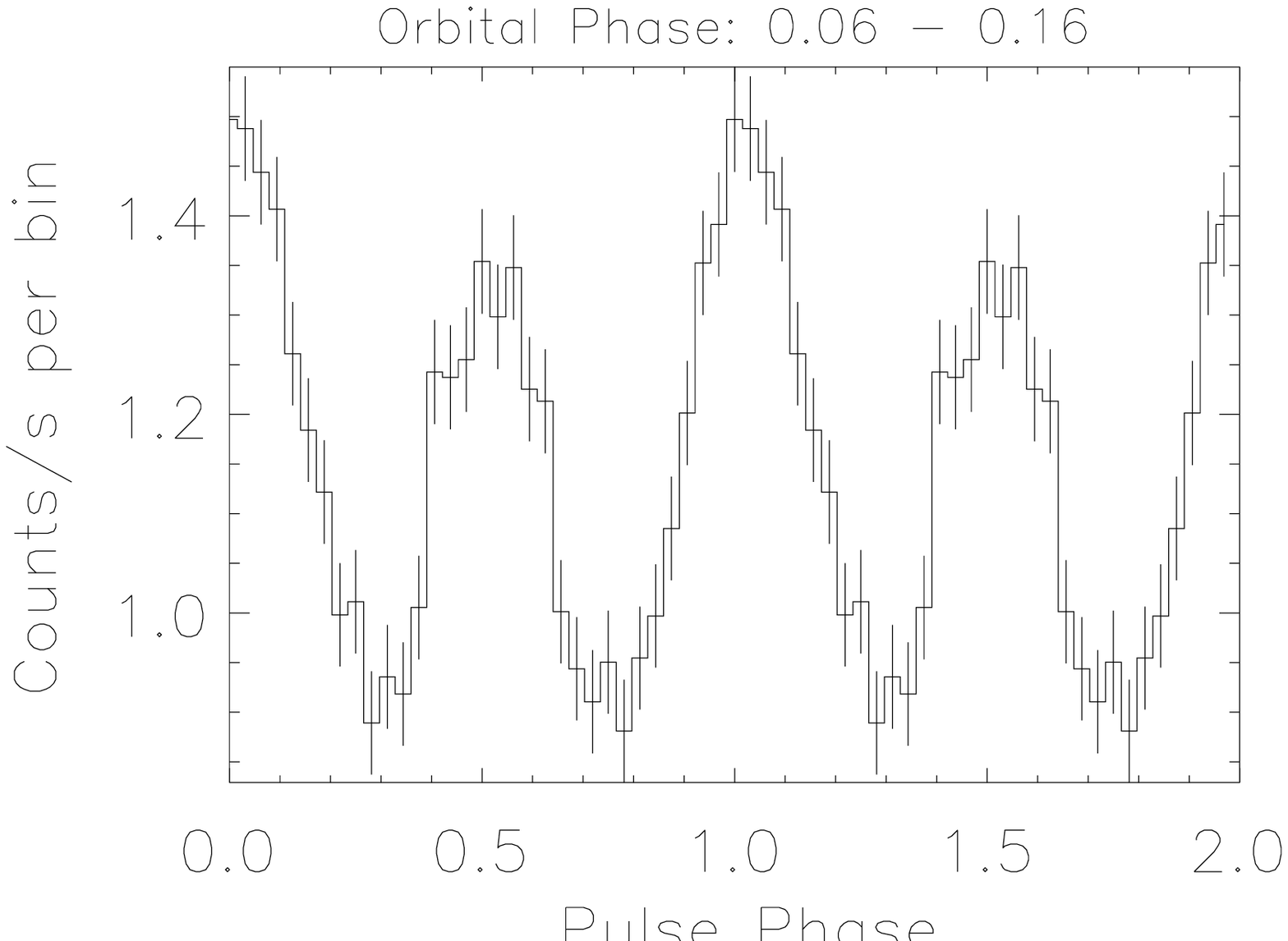}
 }
 \subfigure{
     \includegraphics[width=0.30\textwidth]{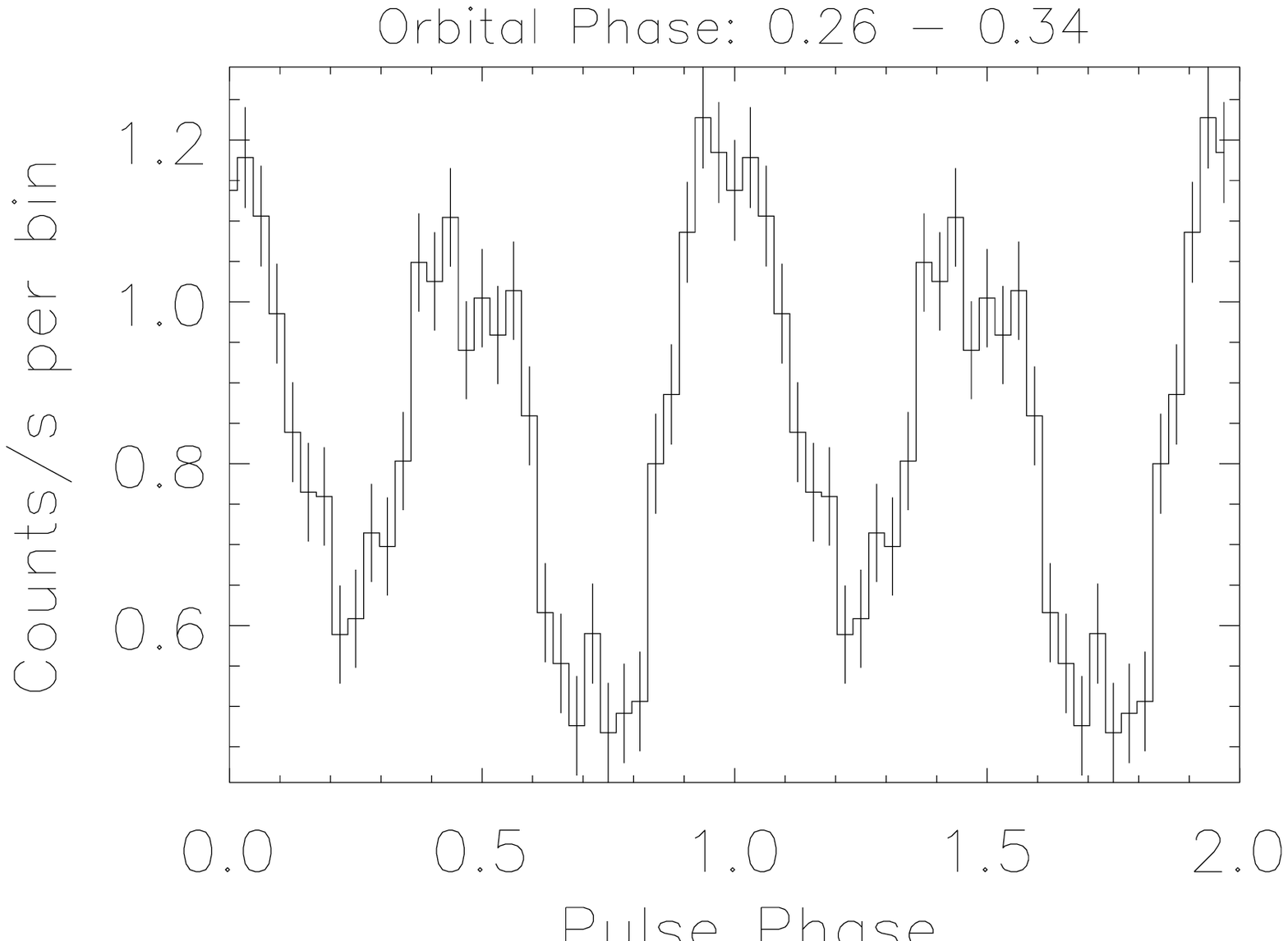}
 }
 \subfigure{
     \includegraphics[width=0.30\textwidth]{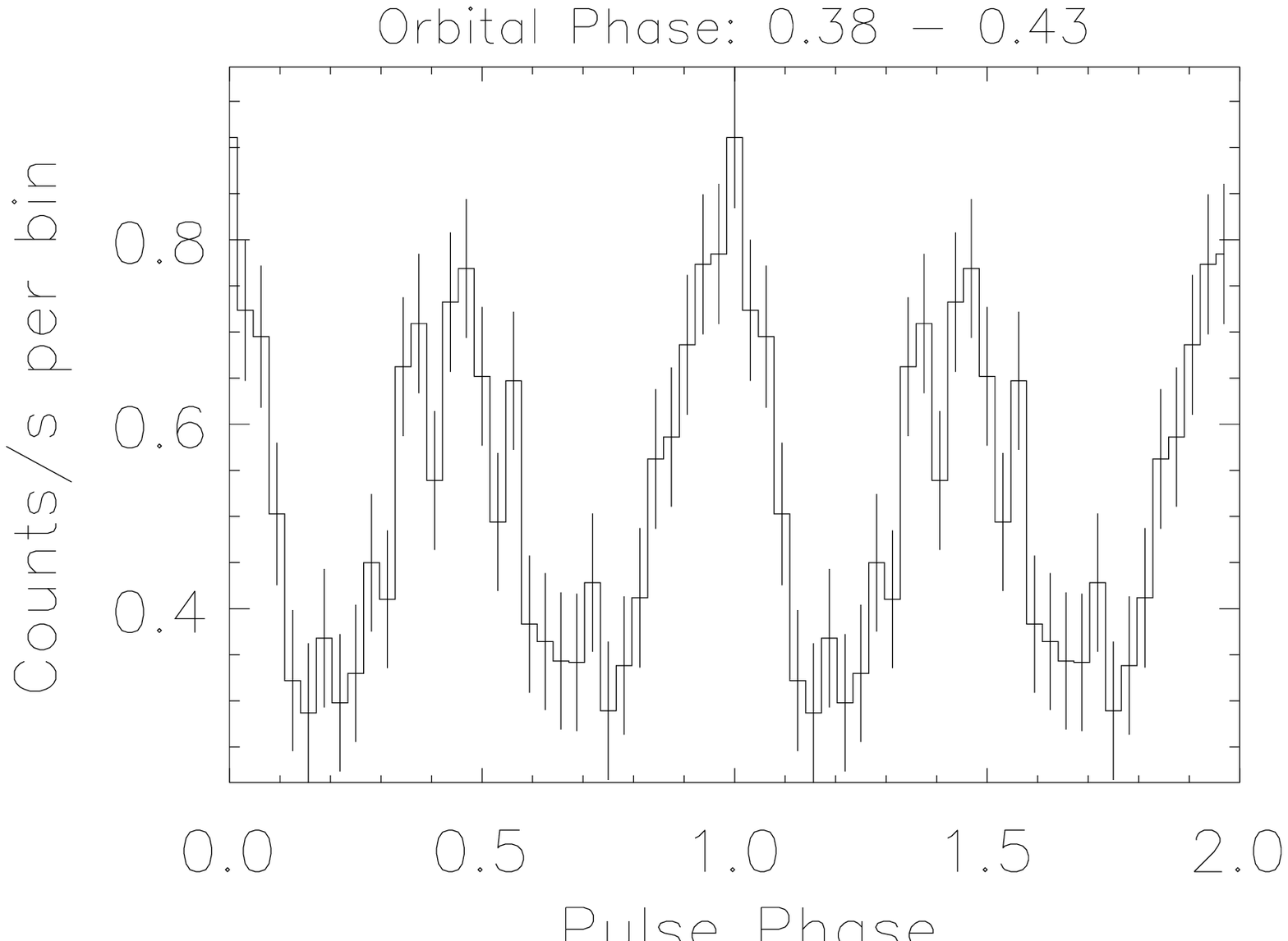}
 }
 \subfigure{
     \includegraphics[width=0.30\textwidth]{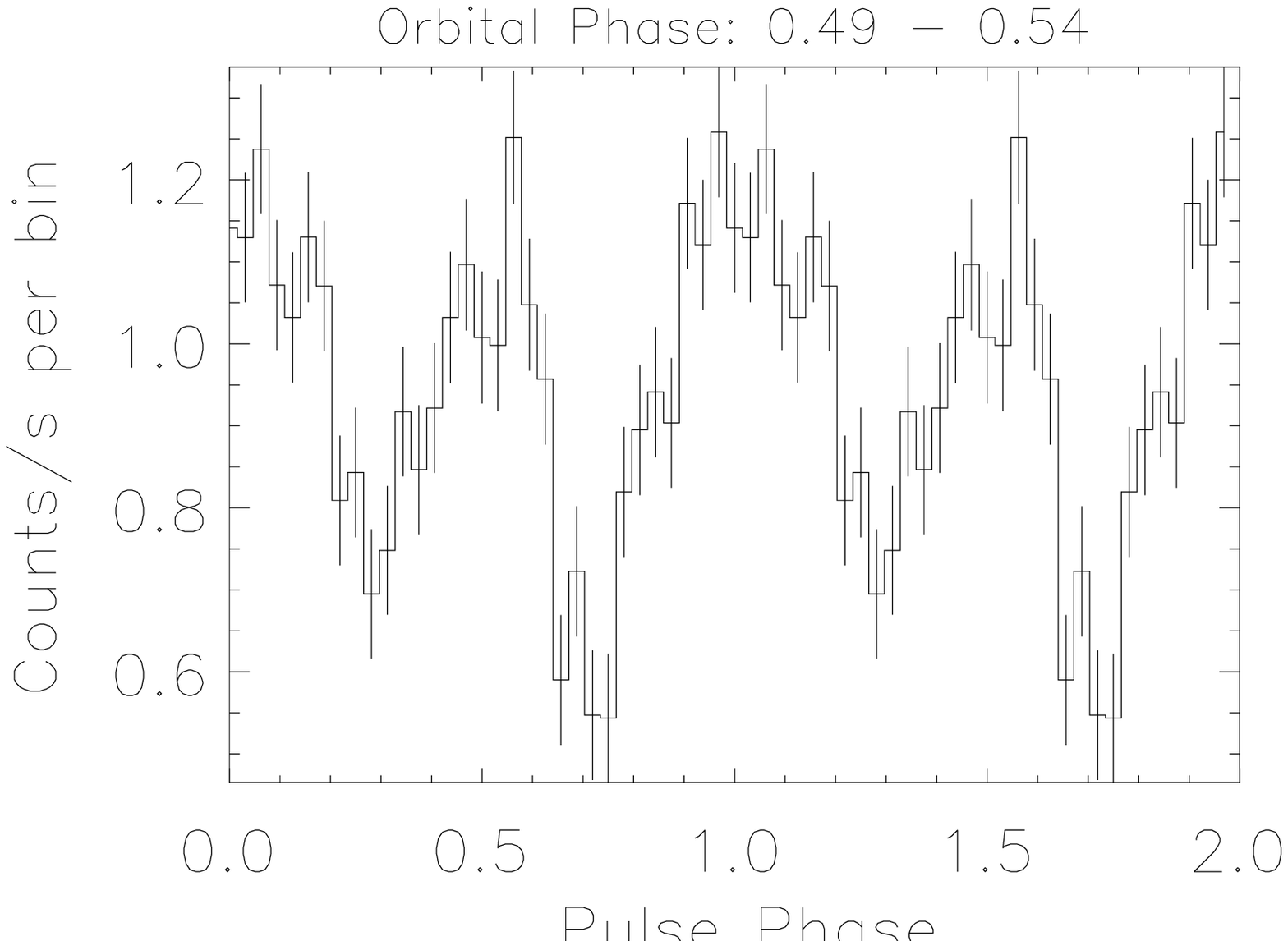}
 }
 \subfigure{
     \includegraphics[width=0.30\textwidth]{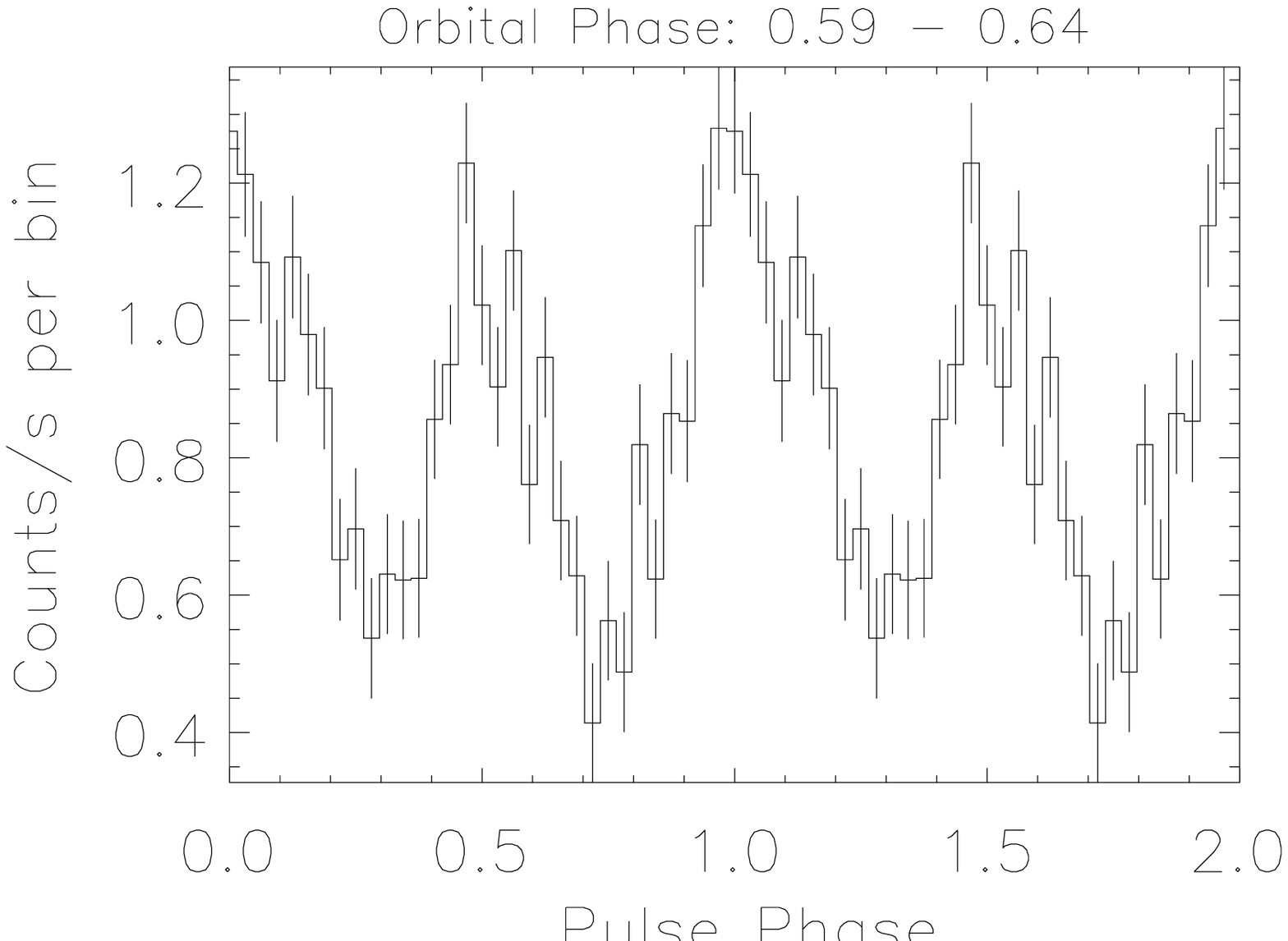}
 }
\subfigure{
     \includegraphics[width=0.30\textwidth]{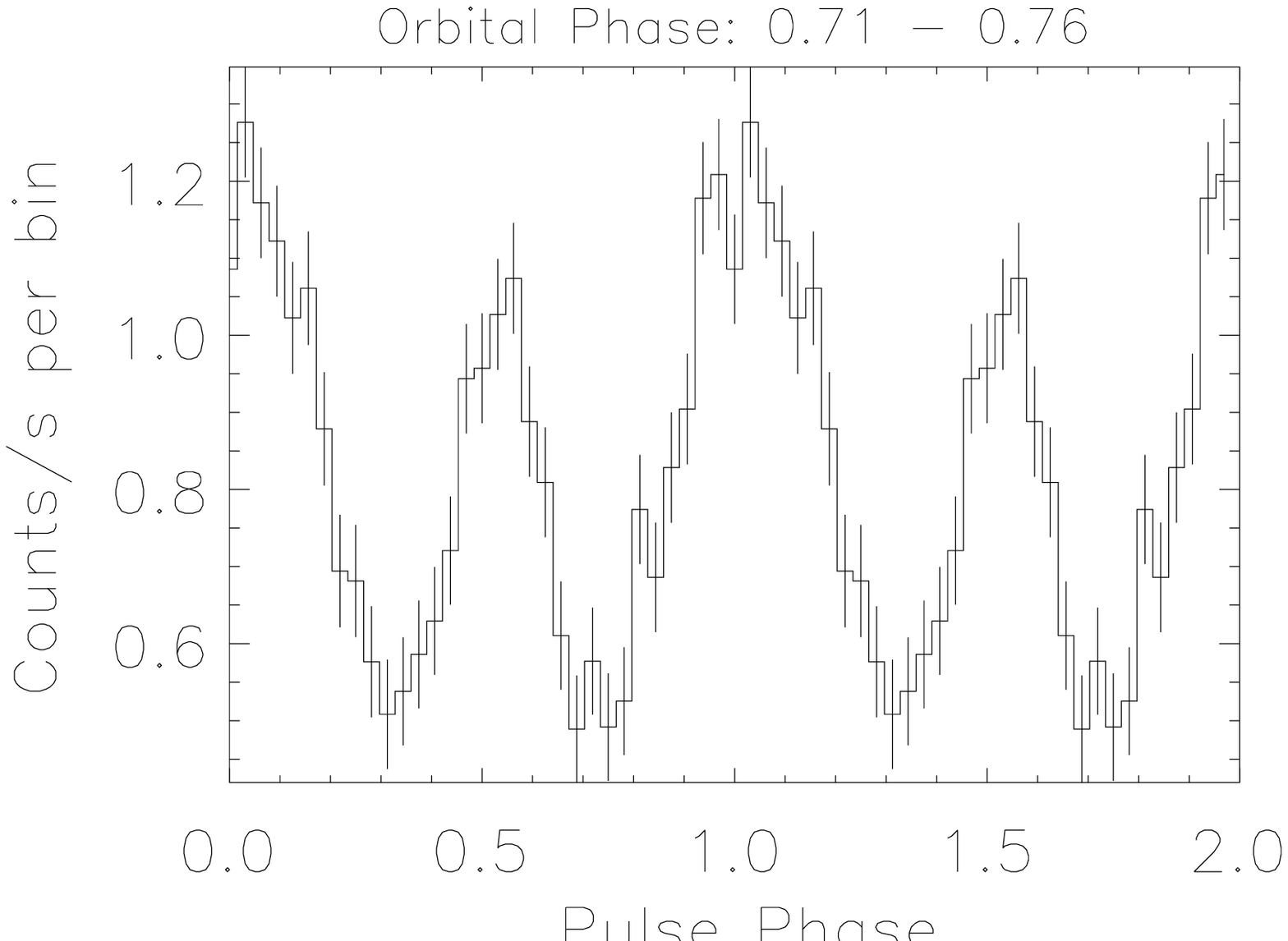}
 }
\subfigure{
     \includegraphics[width=0.30\textwidth]{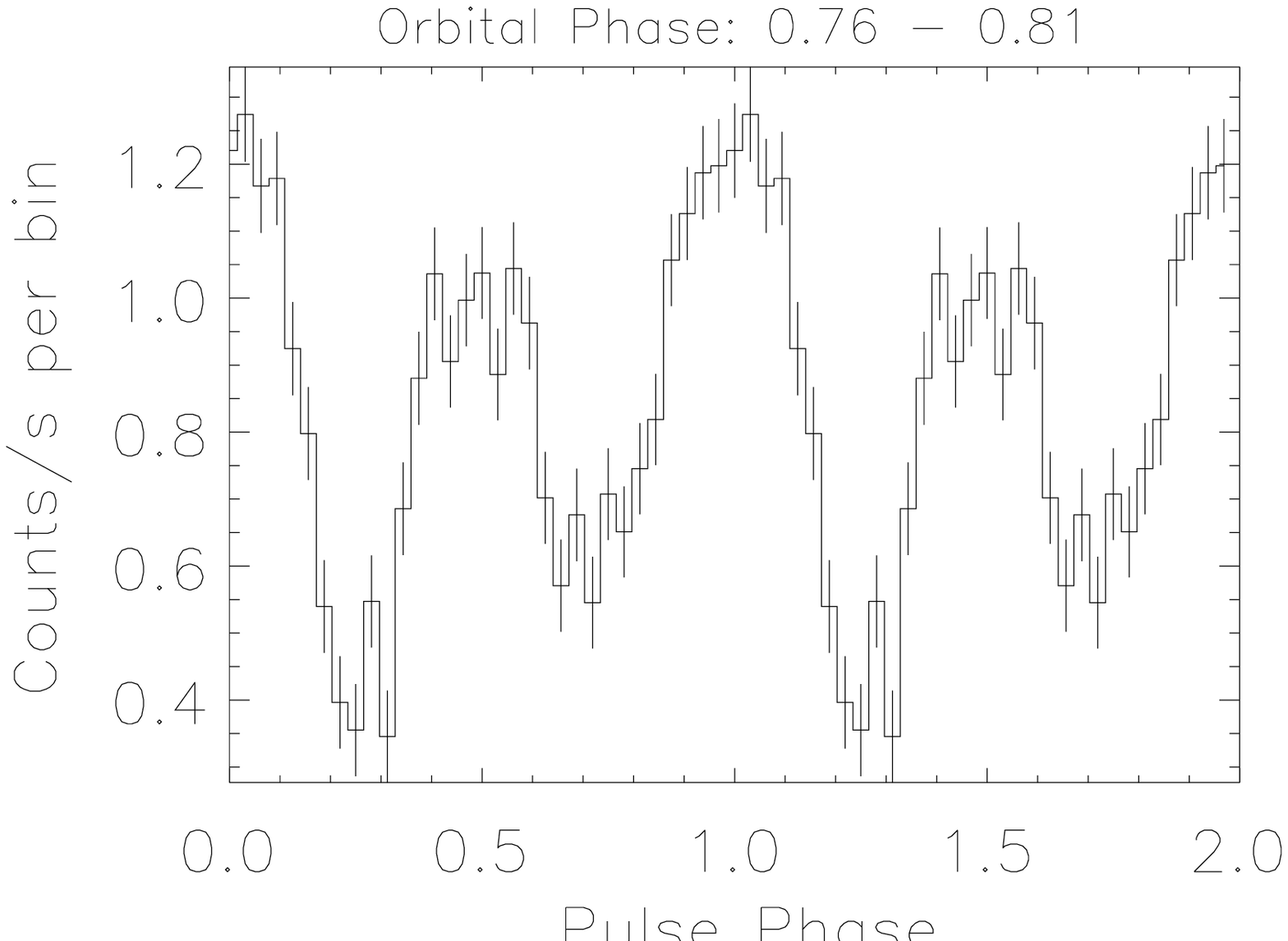}
 }
\subfigure{
     \includegraphics[width=0.30\textwidth]{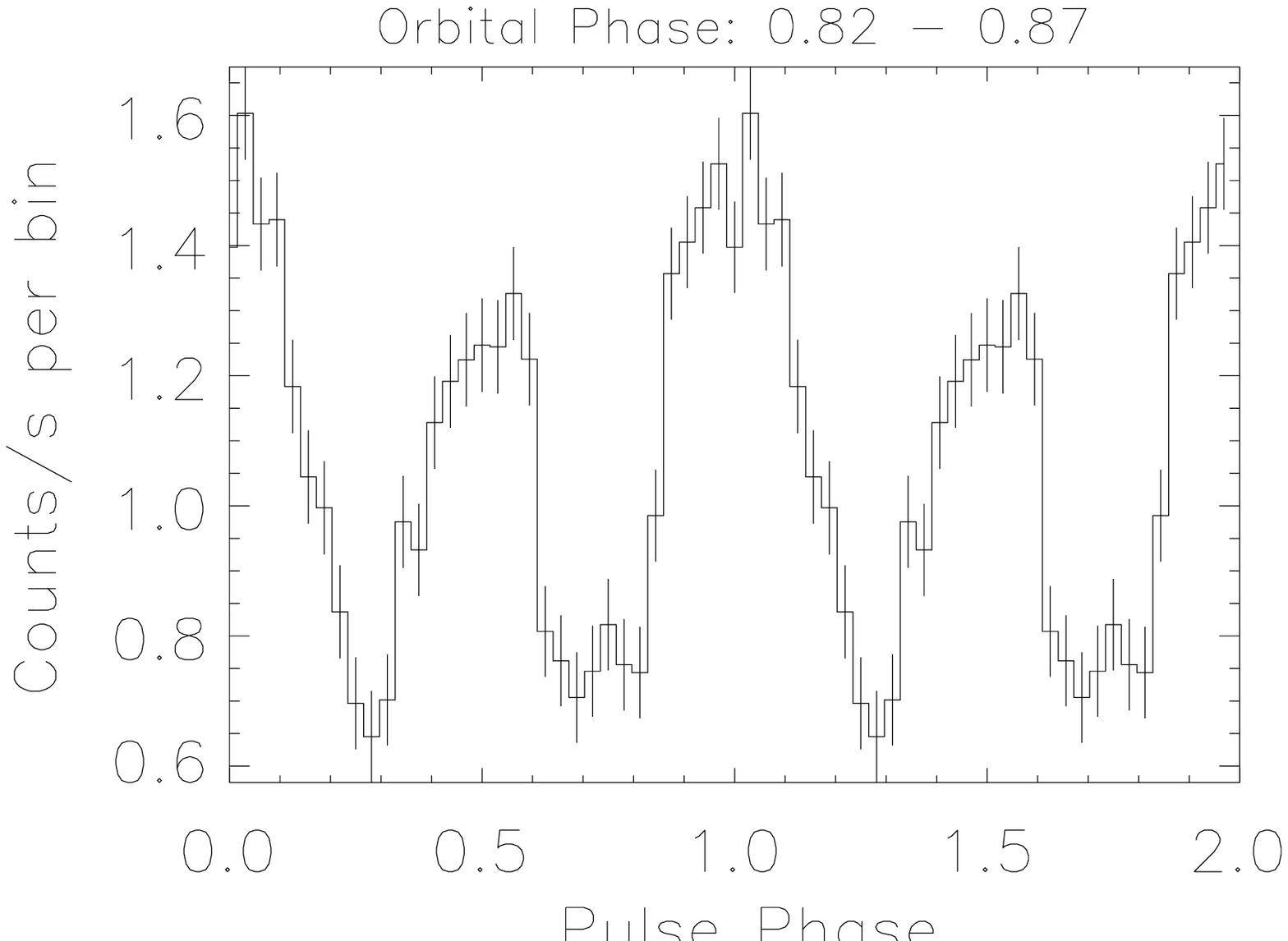}
 }
\subfigure{
     \includegraphics[width=0.30\textwidth]{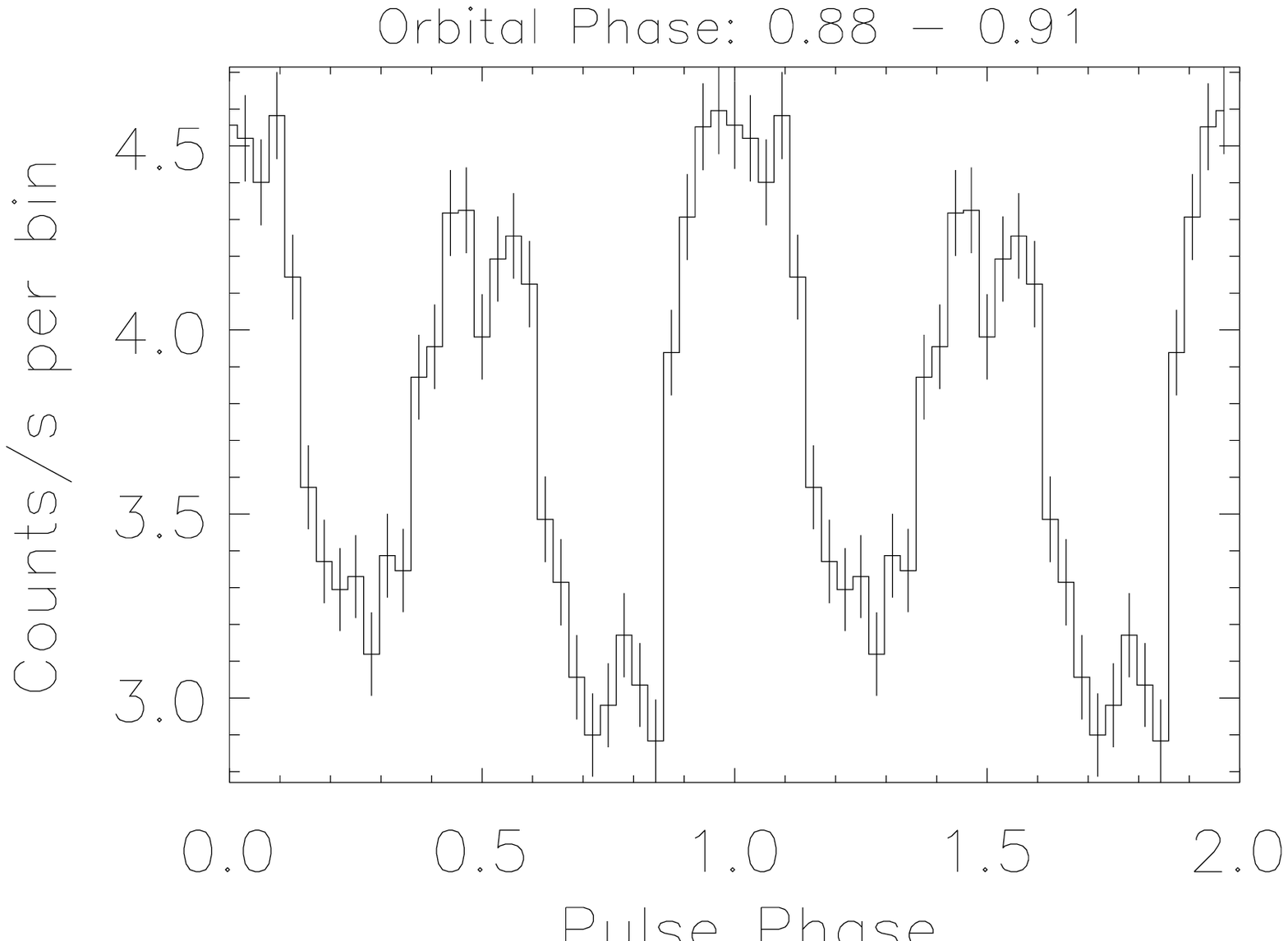}
 }
\subfigure{
     \includegraphics[width=0.30\textwidth]{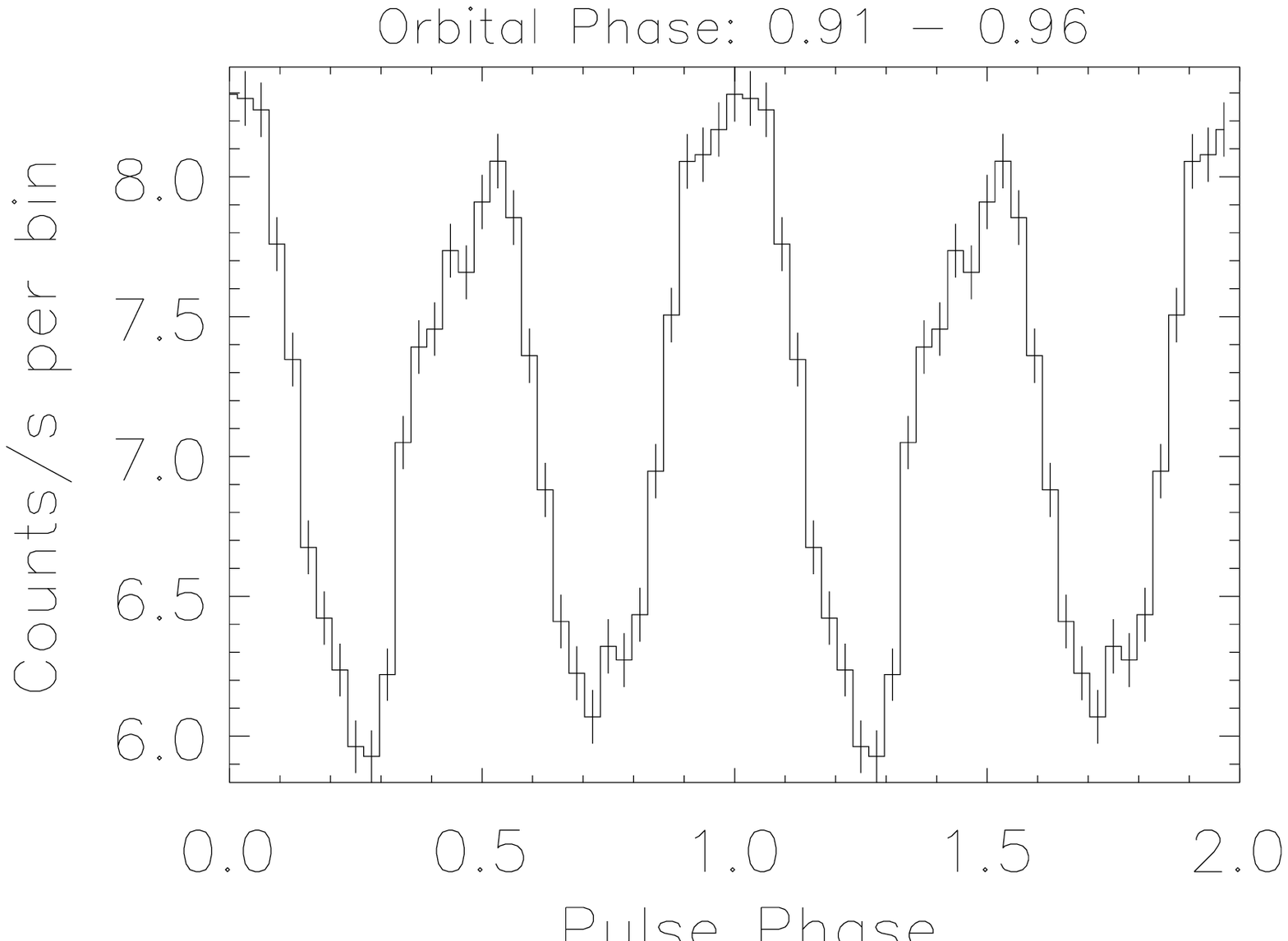}
 }
\subfigure{
     \includegraphics[width=0.30\textwidth]{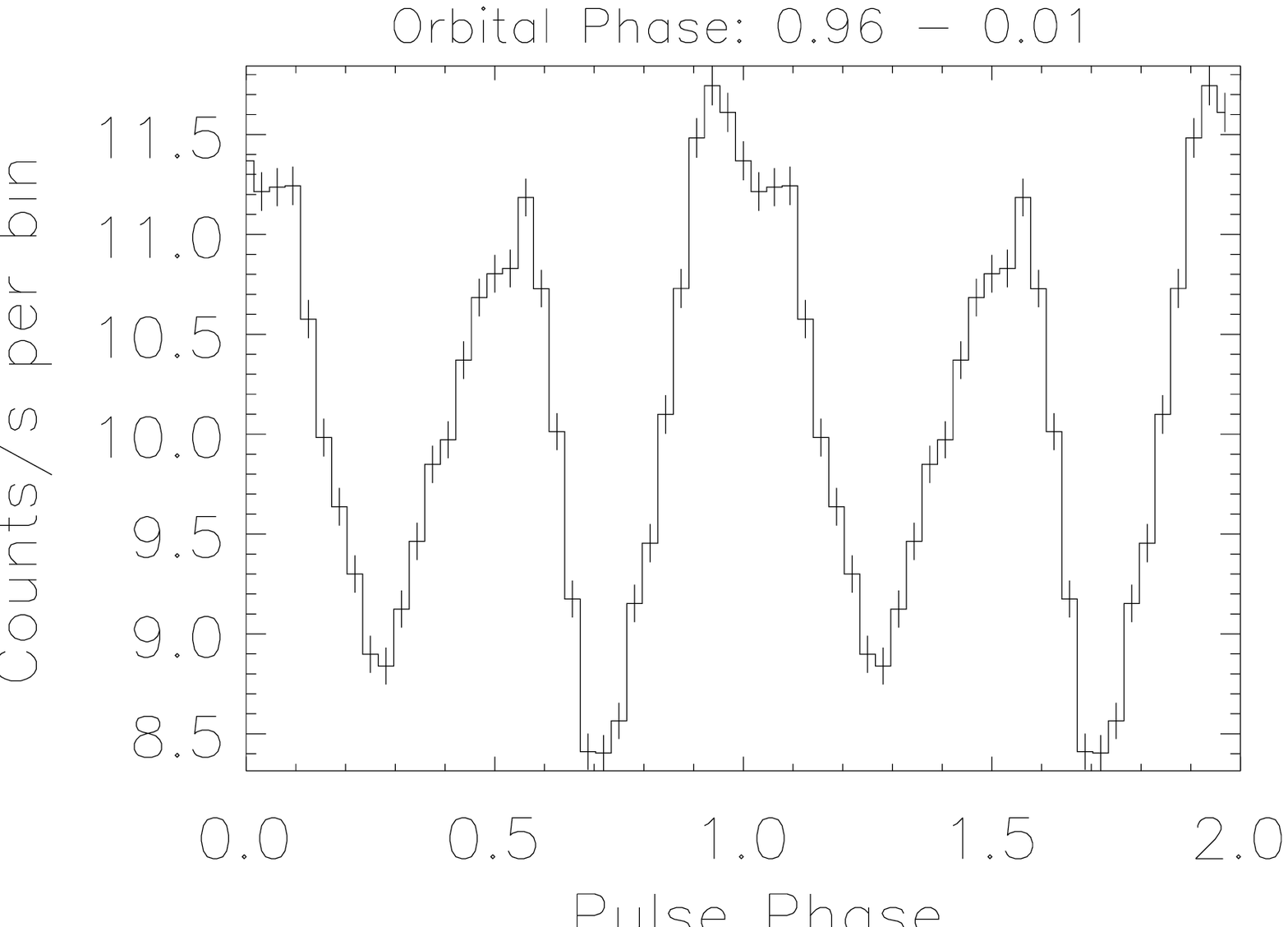}
 }

 \caption{\label{fig:pulse_shape_profile}GX~301-2 pulse shape
   profiles. Each panel represents the pulse shape profile, after the
   background subtraction, in an orbital phase interval as reported on
   top of each figure. See the text for a detailed description of the
   methods used to obtain the pulse shape profiles.} 
\end{figure*}

Due to the erratic behavior of the spin period of the wind-fed
pulsar even on short timescales, as reported by \cite{Nagase1989},
\cite{Bildsten1997}, \cite{Pravdo2001}, and taking into account the
errors on period determination, we were not able to determine the
absolute phasing of the pulse shape profile.
Hence, we cross-correlated each pulse profile with a template
obtained in the orbital phase 0.96--0.01 (lower-right panel in
Figure~\ref{fig:pulse_shape_profile}), which represents the pulse
shape with the highest signal-to-noise.
The resulting phase shifts were then used to build the background
subtracted phase-aligned folded light curves shown in
Figure~\ref{fig:pulse_shape_profile}. 
Despite the different statistical quality of the 
curves, a variability along the orbital cycle can be clearly
identified. The main peaks change in duration, shape and height.
The main peak, at spin phase 0.0, shows a significant double
peak structure only at orbital phases between 0.96 and 0.06, where
the most dramatic flux evolution of the source occurs. Although
these are the orbital phases where the statistical quality is
higher, the insurgence of the minor peak does not appear as an
observational bias: the evolution is gradual and continuous in the
three panels between phases 0.91 and 0.06, all with comparable
quality.

Another variability feature that is worth remarking is the
behavior of the interpulse regions. At spin phase around 0.7 a
small peak appears at orbital phase 0.88-0.91. Its onset seems
already anticipated in the previous orbital phase intervals. 
The other interpulse region, around spin phase 0.3, shows instead an
intensity variability. The source flux in this interval is
significantly higher, with respect to the peaks, in the spin phase
interval between 0.96 and 0.06, the same where the small peak
mentioned above appears.

\subsection{Secular Evolution of the Spin Period}

The SuperAGILE long-term and nearly continuous monitoring of GX~301-2 allowed
us to measure the spin period at different epochs and put it in
the context of the secular timing evolution of the source, as
available from the literature. Figure~\ref{fig:spin_history}(a)
shows the spin period $P$ as a function of the time from the
discovery of the pulsed emission by \cite{White1976} to the
last SuperAGILE observation (MJD 54709.5). This plot confirms that
GX~301-2 is subject to unpredictable torque variations
superimposed on long-term smooth variations of the spin period.
During the last $\sim$30 years the spin period changed passing
from a spin-up state, during which $P$ decreased from $\sim$700.5~s
to $\sim$675.4~s, to a spin-down condition reaching the current
value of $P\sim$687.3~s. The Figure clearly shows the two rapid
spin-up episodes identified by \cite{Koh1997} analyzing {\it CGRO}/BATSE
observations between MJD~48440 and MJD~48463 and between
MJD~49230 and MJD~49245.
Comparing BATSE\footnote{http://gammaray.msfc.nasa.gov/batse/pulsar},
{\it RXTE}/PCA \citep{Kreykenbohm2004}, {\it INTEGRAL} \citep{Doroshenko2008}
and SuperAGILE data in Figure~\ref{fig:spin_history}(a) we find
evidence of at least two other spin-up episodes, occurred between
the observations at MJD~51830 and MJD~52789, and around MJD~54000.

Figure~\ref{fig:spin_history}(b) displays the secular variation of
the GX~301-2 pulse period during the SuperAGILE observations. $P$
varies in the range 685.3--688.0~s in about 410 days.
We find a period evolution consistent with the constant spin down
rate of $4.0 \pm 0.3 \times 10^{-8} \, {\rm s}\, {\rm s}^{-1}$. However, the short 
term variability of the period clearly shows significant timing noise
around the general trend, and the linear fit (dashed line in
Figure~\ref{fig:spin_history}(b)) is indeed a poor modeling of the
period evolution, leading to a reduced $\chi^2$ value of 2.9 (34 dof).

\section{Orbital variability}
\label{sec:orb_lc}

\subsection{SuperAGILE and {\it RXTE}/ASM Folded Light Curves}
\begin{figure}[ht]
 \centering
\subfigure[]{
  \includegraphics[width=0.47\textwidth] {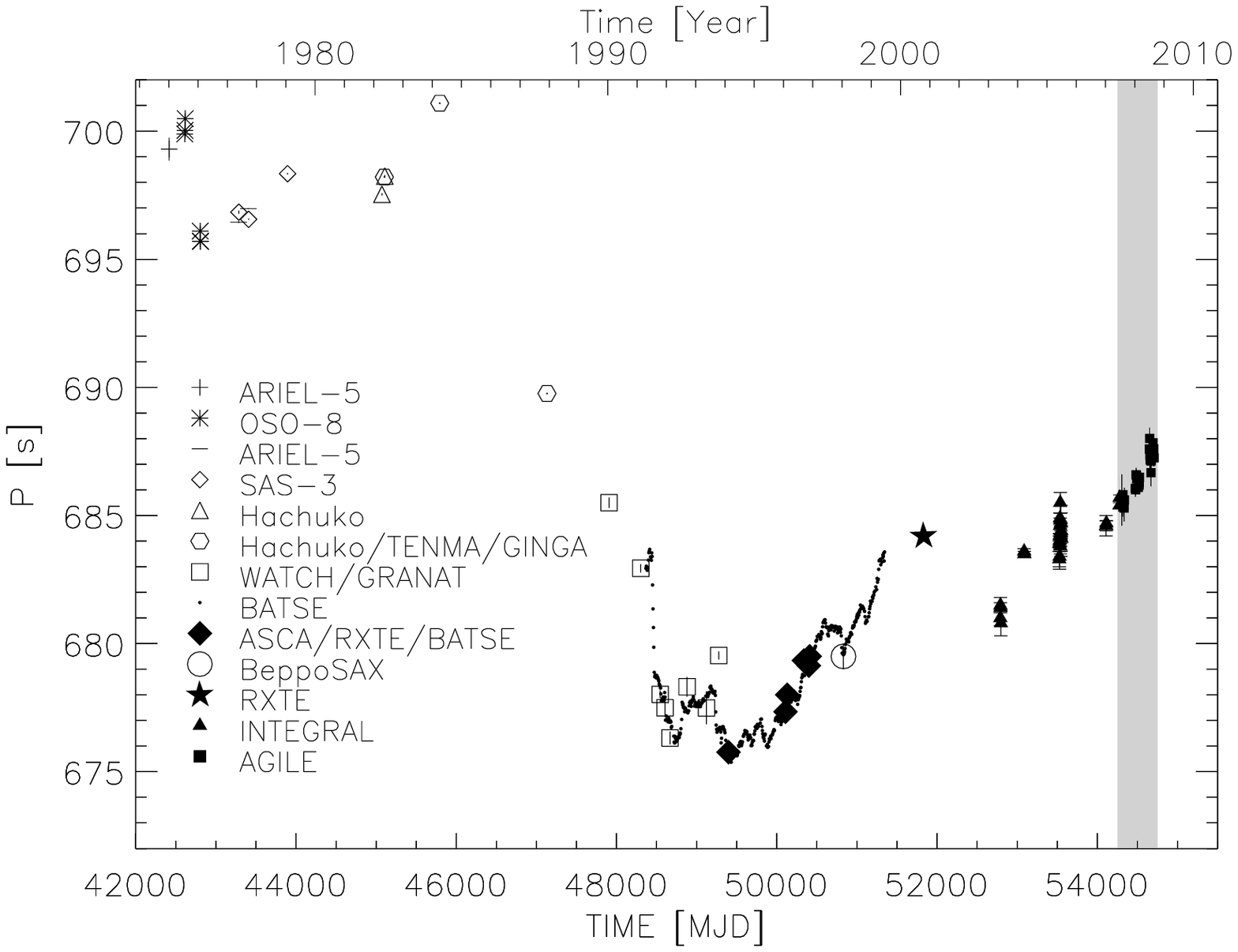}
}
\subfigure[]{
  \includegraphics[width=0.47\textwidth] {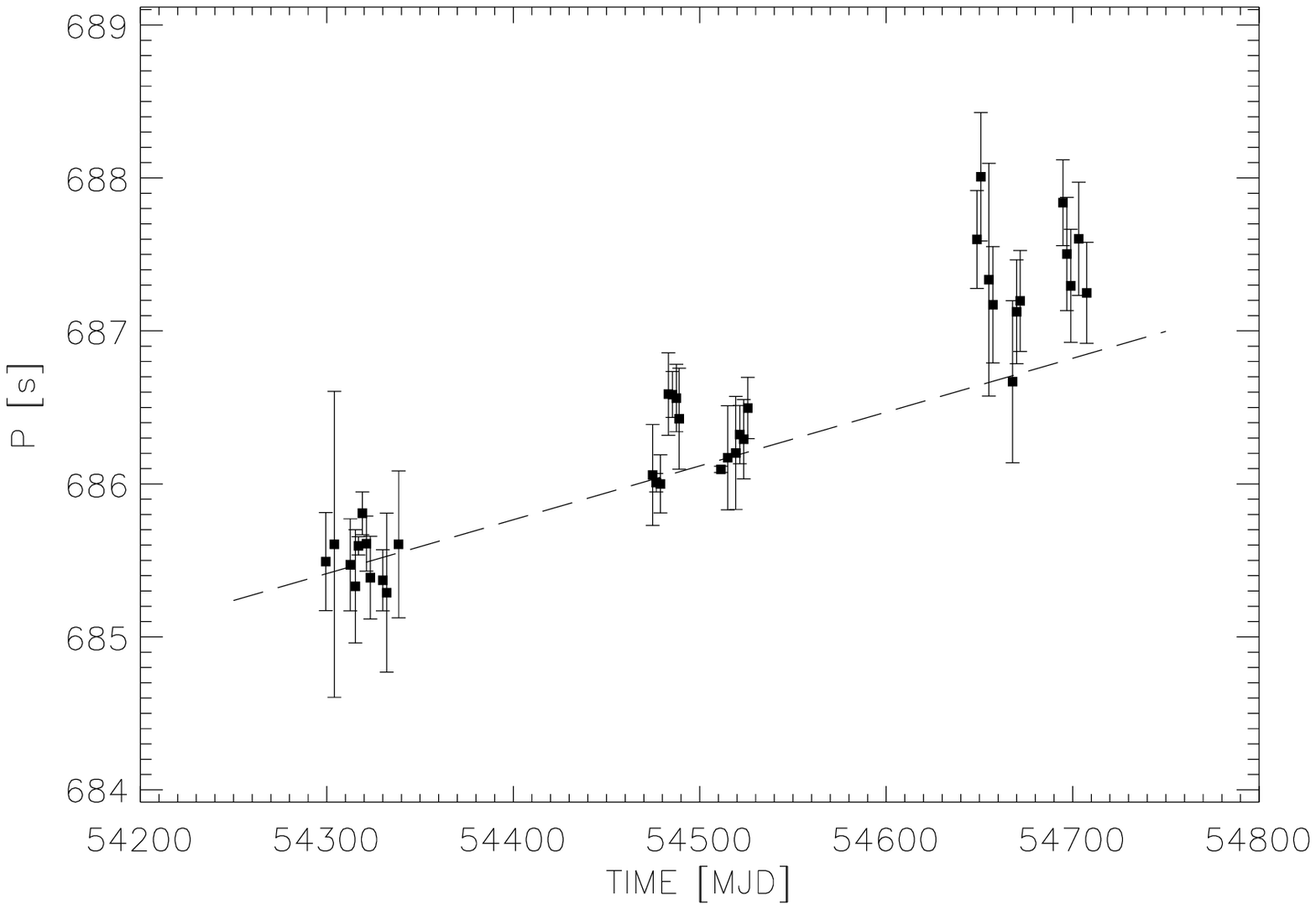}
}
   \caption{\label{fig:spin_history}(a) Secular variations of GX~301-2
     pulse period. Data are from the following references: cross
     symbols \citep{White1976}, asterisks \citep{Swank1976}, dashes
     \citep{White1978}, diamonds \citep{Kelley1980}, triangles
     \citep{Kawai1985}, hexagons \citep{Nagase1989}, squares
     \citep{Chichkov1995}, filled diamonds \citep{Pravdo2001}, stars
     \citep{Kreykenbohm2004}, circles \citep{Labarbera2005}, points
     (http://gammaray.msfc.nasa.gov/batse/pulsar), filled triangles
     \citep{Doroshenko2008}, filled squares in highlighted region
     (SuperAGILE data).  (b) Zoomed view of the SuperAGILE data. The
     dashed line represents the best-fit secular variation of the
     pulse period assuming a linear trend. } 
\end{figure}

\begin{figure}[!h]
 \centering
  \includegraphics[width=0.48\textwidth] {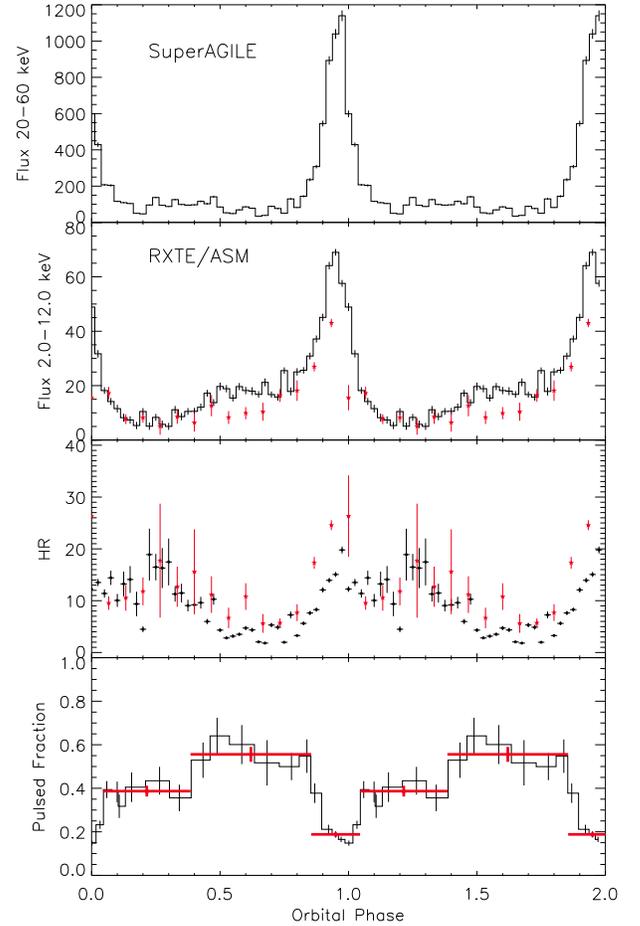}
   \caption{  \label{fig:folded_hr_and_pfrac}Top panel: SuperAGILE
     orbital light curve (20--60~keV) of GX~301-2 folded at orbital 
     period $P_{\mathrm{orb}}$=41.482~d. Second panel: {\it RXTE}/ASM orbital folded
     light  curve (2--12~keV) of GX~301-2 obtained using the whole
     ASM dataset from MJD~50087.285 to MJD~54720.745 (black solid 
   line) and only the observations contemporaneous to SuperAGILE (red
   stars). Both the folded light curves are given in mCrab
   units. Third panel: Hardness ratio (20--60~keV to 2--12~keV)
   calculated using the whole SuperAGILE and ASM dataset (filled
   circle) and the contemporaneous observations only (red
   stars). Bottom panel: pulsed fraction ($P_{\mathrm{frac}}$)  as a function
   of the orbital phase.  Thick points represent the mean pulsed
   fraction values in the orbital phases 0.05--0.39, 0.39--0.86 and
   0.86--0.05.  The sharp $P_{\mathrm{frac}}$ minimum which corresponds with
   the pre-periastron flare is clearly visible.} 
\end{figure}

Using  all the SuperAGILE data collected during the observations
listed in Table~\ref{tab:pointings}, we obtained the complete,
folded orbital light curve of GX~301-2 shown in the top panel of
Figure~\ref{fig:folded_hr_and_pfrac}. The orbital parameters ($P_{\mathrm{orb}}$,
$T_{0}$) used in the folding procedure were the same used for the
correction for the time delay discussed in
Section~\ref{sec:timing}. The difference in the
orbital solution found by \cite{Doroshenko2008} and by \cite{Koh1997}
results in a 5\% phase shift of the PP flare in the orbital
folding, while the shape of the folded light curve does not change
noticeably. In order to obtain a folded light curve in the soft
X-ray band we used public data from the All Sky Monitor onboard
{\it RXTE} \citep{Levine1996}. ASM data for GX~301-2 cover the time
span from MJD~50087.285 to MJD~54720.745 and consist of source
count rates for each ASM dwell (1 dwell = 90~s observation) in the
band 2~keV-12~keV. The dwell-by-dwell data were filtered
according to the criteria reported on {\it RXTE}/ASM web
site\footnote{http://heasarc.gsfc.nasa.gov/docs/xte/SOF/asmlc.html}
and then a light curve was built from the cleaned data. The ASM
folded light curve for GX~301-2 is displayed in the second panel
of Figure~\ref{fig:folded_hr_and_pfrac} (solid line).

Both the SuperAGILE and the ASM light curves are given in units of
mCrab. We used the standard value of 1~Crab~=~75~counts~s$^{-1}$ as a
conversion factor for the ASM count rates, while the SuperAGILE
normalization is 0.15~counts~cm$^{-2}$~s$^{-1}$. The
two folded orbital light curves clearly show the PP flaring activity,
with an intensity peak observed at orbital phase between $\sim$0.9
and $\sim$0.0 and a full width at half maximum of about 0.1 in
phase. Fitting the PP flare by using a Gaussian function plus a
constant term, we obtained a value of orbital phase of 0.954$\pm$0.005
for the flare position, with a FWHM value of 0.109.
The remaining orbital interval exhibits a different behavior in the
hard and soft X-ray data: in the 20-60~keV energy band a nearly
constant persistent emission is detected between phases 0.1 and 0.8,
while in the same phase interval the soft X-ray data show a broad
hump, covering phases between 0.5 and 1.0.

\subsection[]{Variability of the pulsed fraction}
\label{sec:pulsed_frac}

The variability in the shape of the pulse as a function of the
orbital phase shown in Figure~\ref{fig:pulse_shape_profile}
motivated us to study the pulsed fraction across the binary
orbital cycle, to determine its possible variability and
correlation with other source parameters, such as flux and hardness.
We operationally define the pulsed fraction ($P_{\mathrm{frac}}$)
as the ratio between the pulsed counts, as derived directly from 
folding the event list, including photons from the background and
from other sources in the FoV, and the total counts determined by the
imaging procedure, including the counts from GX~301-2 only. 
We performed the analysis of the pulsed fraction as a function of the
pulsar orbital phase for all the available SuperAGILE
observations during which the pulsation was detected. The results are
shown in the bottom panel of Figure~\ref{fig:folded_hr_and_pfrac},
where $P_{\mathrm{frac}}$ is plotted as a function of the orbital phase.
\begin{table}[t]
  \begin{center}
    \caption{ \label{tab:pp_hr} Mean Hardness Ratio of the PP Flares}
    \begin{tabular}{ccccc}
      \small{MJD Start} & \small{MJD Stop} & \small{Phase Start} & \small{Phase Stop} & \small{HR} \\
      \tableline\tableline
      54315.84 & 54319.37 & 0.91 & 0.99 & 10.4 $\pm$ 0.6\\
      54481.77 & 54485.30 & 0.91 & 0.99 & 17.5 $\pm$ 1.2\\
      54523.25 & 54526.78 & 0.91 & 0.99 & 21.8 $\pm$ 1.7\\
      54647.70 & 54652.22 & 0.91 & 0.99 & 14.0 $\pm$ 0.9\\
      \tableline
    \end{tabular}
  \end{center}
\end{table}

$P_{\mathrm{frac}}$ exhibits large variations across the binary cycle,
passing from a minimum of $P_{\mathrm{frac}}=0.15\pm 0.02$ at orbital phase
$\sim$0.0 to a maximum of $P_{\mathrm{frac}}=0.64\pm 0.12$ at orbital phase
$\sim$0.5. With the aim of quantifying the significance of the
variability of the pulsed fraction, we took the average of
$P_{\mathrm{frac}}$ in the phase intervals where it appears consistent with
a constant value, that is 0.05--0.39, 0.39--0.86 and
0.86--0.05. Thick red points in the bottom panel of
Figure~\ref{fig:folded_hr_and_pfrac} represent the mean $P_{\mathrm{frac}}$
values calculated during the three 
selected orbital intervals. The values in the three phase 
intervals are 0.188$\pm$0.009, 0.387$\pm$0.025 and 0.556$\pm$0.034, ordered
by phase. This analysis shows that it is possible to identify
three different ``states'' in the pulsed fraction as a function of the
NS orbital motion, with the intermediate one occurring just after the
PP flares and lasting approximately up to the phase corresponding 
to the superior conjunction of the binary system. 

\begin{figure}[!t]
 \centering

 \subfigure[]{
  \includegraphics[width=0.48\textwidth]{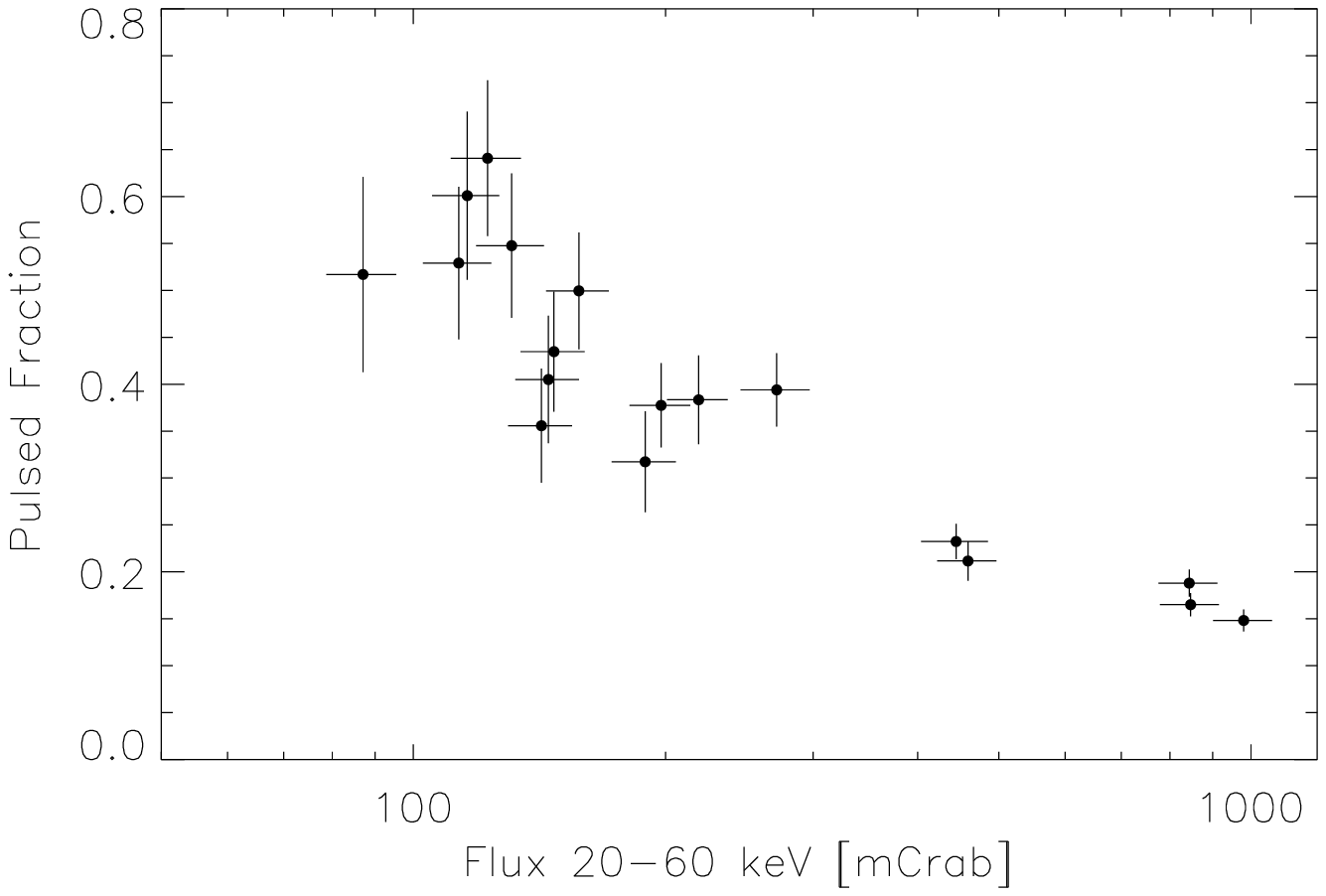}
 }
 \subfigure[]{
  \includegraphics[width=0.48\textwidth]{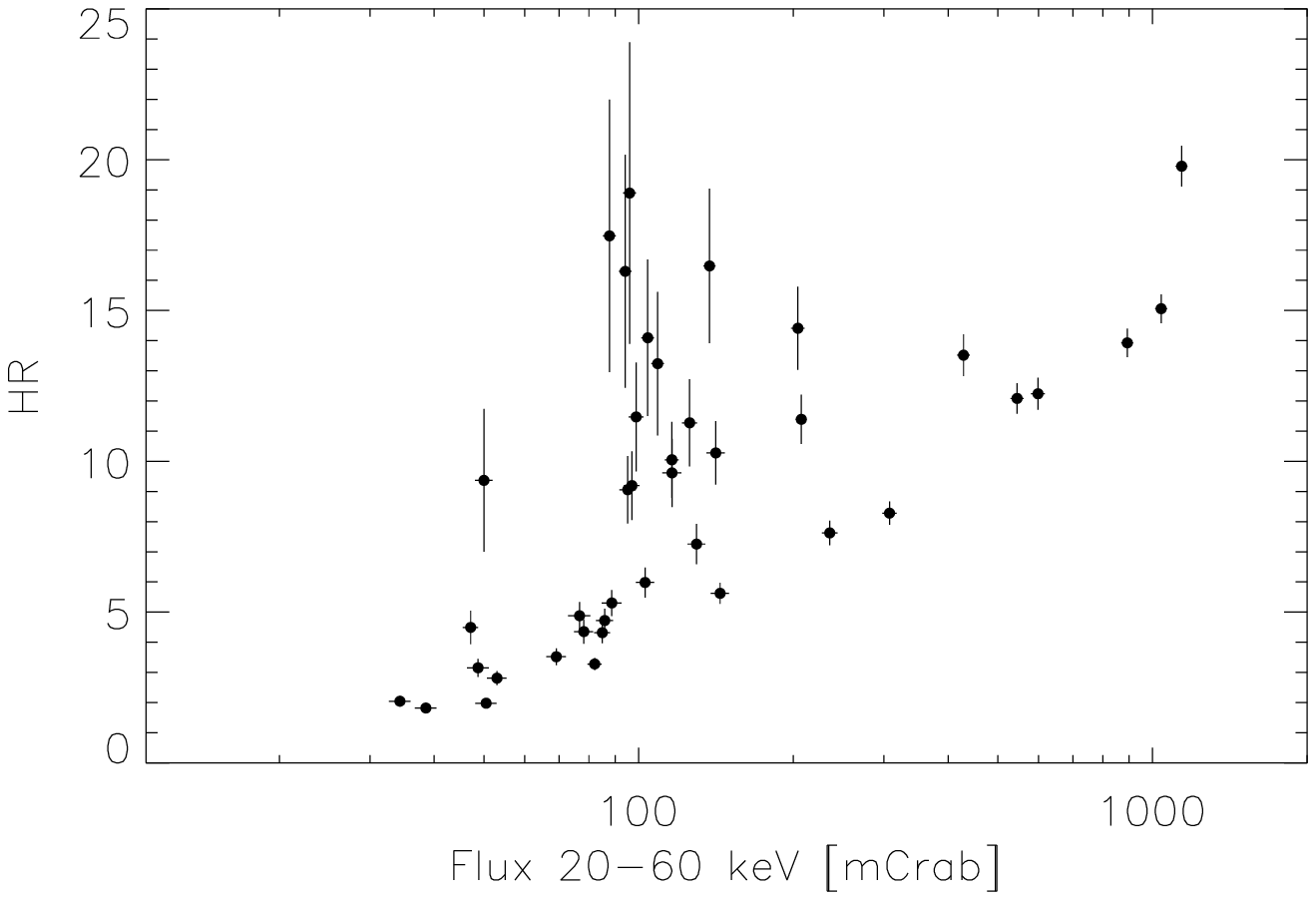}
 }
 \subfigure[]{
  \includegraphics[width=0.48\textwidth]{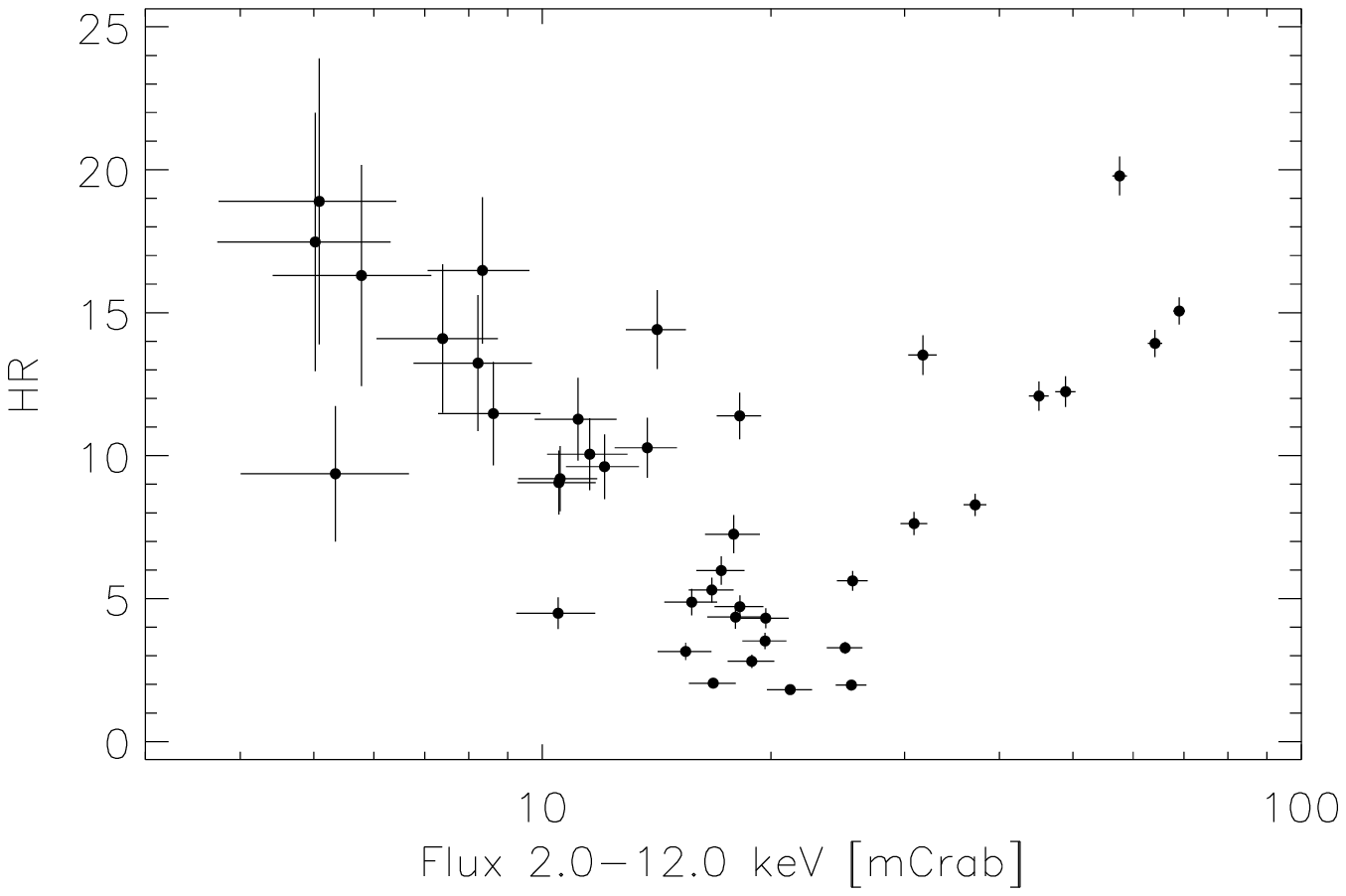}
 }   \caption{  \label{fig:pulsed_frac}(a) $P_{\mathrm{frac}}$ as a function of
   the source intensity. The clear anti-correlation between pulsed
   fraction and source flux is ascribed to the different amount of
   matter which accumulates near and accretes onto the neutron
   star. (b) and (c) Hardness ratio as a function of the source flux
   in the 20--60~keV (b) and 2--12~keV (c) energy bands.}
\end{figure}

The bottom panel in Figure~\ref{fig:folded_hr_and_pfrac} shows that the pulsed fraction reaches
its minimum at the phase when the large flare occurs, that
is where the largest variations in the pulse shape are detected.
We then investigated the possible dependence of $P_{\mathrm{frac}}$ on
the source flux. The results are shown in
Figure~\ref{fig:pulsed_frac}(a), where the value of the pulsed fraction
against the source flux is displayed, independently of the orbital or
spin phase of the source. A clear anti-correlation between these two
quantities is found, confirming it as a general feature and not only
an event occurring during the PP periodic flare. 
The plot shows that the pulsed fraction decreases rapidly at increasing fluxes, 
reaching a plateau for large fluxes corresponding to a pulsed
fraction of 15\%, approximately a factor 4 smaller than the maximum
value.
A similar behavior was also recently reported by  \cite{Lutovinov2009}
using {\it INTEGRAL}/IBIS data.

\subsection{Hardness Ratio Versus Orbital Phase}

Given the different orbital flux modulation at soft and hard X-rays
(Figure~\ref{fig:folded_hr_and_pfrac}), we 
investigated the period-averaged time behavior of the source
hardness along the orbital light curve. To this aim we computed an
hardness ratio (HR) by using the flux measurements by SuperAGILE
and ASM, that is between the energy ranges 20--60~keV and 2--12~keV. 
The third panel in Figure~\ref{fig:folded_hr_and_pfrac} shows the
result. Filled circles indicate the orbital phase-resolved
hardness ratio obtained by using the complete temporal data set
from both experiments. Moreover, star-like symbols provide the same
quantity, but computed after selecting the ASM time intervals
coeval with the SuperAGILE data sets (the ASM light curve with
this selection is shown using the same symbols in the second
panel of the same figure).

The time profile of the HR computed using both time selections
indicates a clear modulation of the source spectral hardness as a
function of the orbital phase, where we identify the following
features. A spectral hardening is detected during the
pre-periastron flare: the hard X-ray flux rises faster than the
soft X-rays, and lasts longer. Then, when the soft X-ray flux drops
at its minimum orbital level, around phase 0.2, the hard X-rays
hold their intensity level, indicating an increasing hardness of
the emitted radiation. Subsequently, the rise of the soft X-ray
flux after phase 0.5 (the apastron) is not accompanied by a
corresponding increase at hard X-rays, indicating a softer
emission in this part of the orbit. We note that using the
simultaneous data only (stars in the HR plot), the above features
are confirmed and, within the limits of the lower statistical
quality, emphasized. Actually, the source spectral hardness
changes also from one binary orbit to another. In
Table~\ref{tab:pp_hr} we provide the HR values computed in the
orbital phases including the PP flare in the four available
SuperAGILE observations. The value of the HR changes up to a
factor of 2.

\section{Discussion}
\label{sec:discussion}

The pointing strategy of the {\it AGILE} mission offered SuperAGILE the
opportunity of studying the temporal properties of GX~301-2 in the
hard X-rays over repeated $\sim$month-long stretches of time,
covering six 41.5~days binary orbits at different phases. We observed
a modulation of the hard X-ray flux along the orbit in general
agreement with what reported over the past two decades by previous
experiments in soft and hard X-rays. 
Fitting the source spectrum in the 20--60~keV energy band with a
cut-off power-law with photon index fixed to $\Gamma$=1.05
\citep{Labarbera2005}, we obtain 
$E_{\mathrm{cut}}$=24.4$^{+1.2}_{-1.4}$~keV and
$E_{\mathrm{fold}}$=7.8$^{+0.7}_{-0.7}$~keV for the PP flare occurred
around MJD~54316. Using these parameters, and assuming a source
distance of 3~kpc \citep{Kaper2006}, we convert the SuperAGILE counts
to physical units obtaining a luminosity of $\sim$2.3$\times
10^{37}$~erg~s$^{-1}$. Under the same assumptions, the luminosity
during the quiescent orbital phases is approximately one order of
magnitude smaller.

The only complete folded orbital light curve of GX~301-2 at energies
above 20~keV available in the literature, to which we can compare the
SuperAGILE results, was built from the data of the BATSE experiment,
using the Earth occultation technique \citep{Pravdo1995, Pravdo2001} 
or the pulsed flux measurements \citep{Koh1997}. The SuperAGILE
data show significant differences. In contrast to the BATSE
observations in the same energy range \citep{Pravdo2001}, the
SuperAGILE data do not show any secondary peak in the folded light
curve near the passage at the apastron (orbital phase $\sim$0.5).
This second peak was indeed weakly detected in the BATSE
measurements of the total source flux \citep{Pravdo2001},
while it was outstanding in the pulsed flux only \citep{Koh1997}. The
different shape of the orbital light curve of the total flux as
measured by BATSE and SuperAGILE may be ascribed to
intrinsic time variability between the different orbits and/or
epochs of the observations (1991--1997 for BATSE and 2007--2008 for
SuperAGILE) and/or to the different effective bandpasses of the two
experiments (harder for BATSE than for SuperAGILE despite their
similar nominal energy range). 
Actually, hints toward time variability near the apastron phase
were already provided by the epoch-resolved BATSE folded light
curves reported by \cite{Pravdo2001}, where only two out of
four curves show a broad hump around phase 0.5, although the
statistical quality of the data does not allow for a definite
statement. Also the recent {\it Swift}/BAT data of three orbital cycles
reported by \cite{Pestalozzi2009} show bumps around phase $\sim$0.5 in
two contiguous orbital cycles around 2008 December (unfortunately
these time periods were not covered by SA observations).

Instead, the pronounced bump around the apastron phase found by
\cite{Koh1997} in the pulsed flux appears consistent with our
results, when the behavior of the pulsed fraction along the
binary orbit is considered. As we show in
Figure~\ref{fig:folded_hr_and_pfrac}, we found a large orbital
variation of the pulsed fraction: within the statistical accuracy of
our data, the pulsed fraction exhibits a step-like behavior along the
orbit. The minimum value is reached at the periastron, while a broad
maximum, nearly 3 times as high, is reached at phases between 0.4 and
0.9. This implies that the pulsed flux increases in the phase range
corresponding to the apastron, coherently with what was found in the BATSE
pulsed data.

Interestingly, the same orbital phases where the pulsed fraction
is highest are also those where the emitted radiation is softest.
In fact, we analyzed the behavior of the source hardness along
the orbit by using the SuperAGILE data in combination with the
public {\it RXTE}/ASM data. We found that the radiation emitted by the
source is harder near the periastron and at phases around 0.2--0.3,
while it becomes significantly softer at phases 0.5--0.8
(Figure~\ref{fig:folded_hr_and_pfrac}).
The hardening around phase 0.2--0.3 was also detected by
\cite{Pravdo2001} through an analysis similar to ours, using the BATSE 
and ASM data, but they did not report evidence for the hardening
near the periastron, that is instead the main feature in our data.
More in general, our data suggest an anti-correlation between the
hardness and the pulsed fraction at hard X-rays (as well as
between the pulsed fraction and the flux,
Figure~\ref{fig:pulsed_frac}(a)).
\begin{figure*}[!t]
 \centering
 \subfigure[]{
     \includegraphics[height=0.3\textheight]{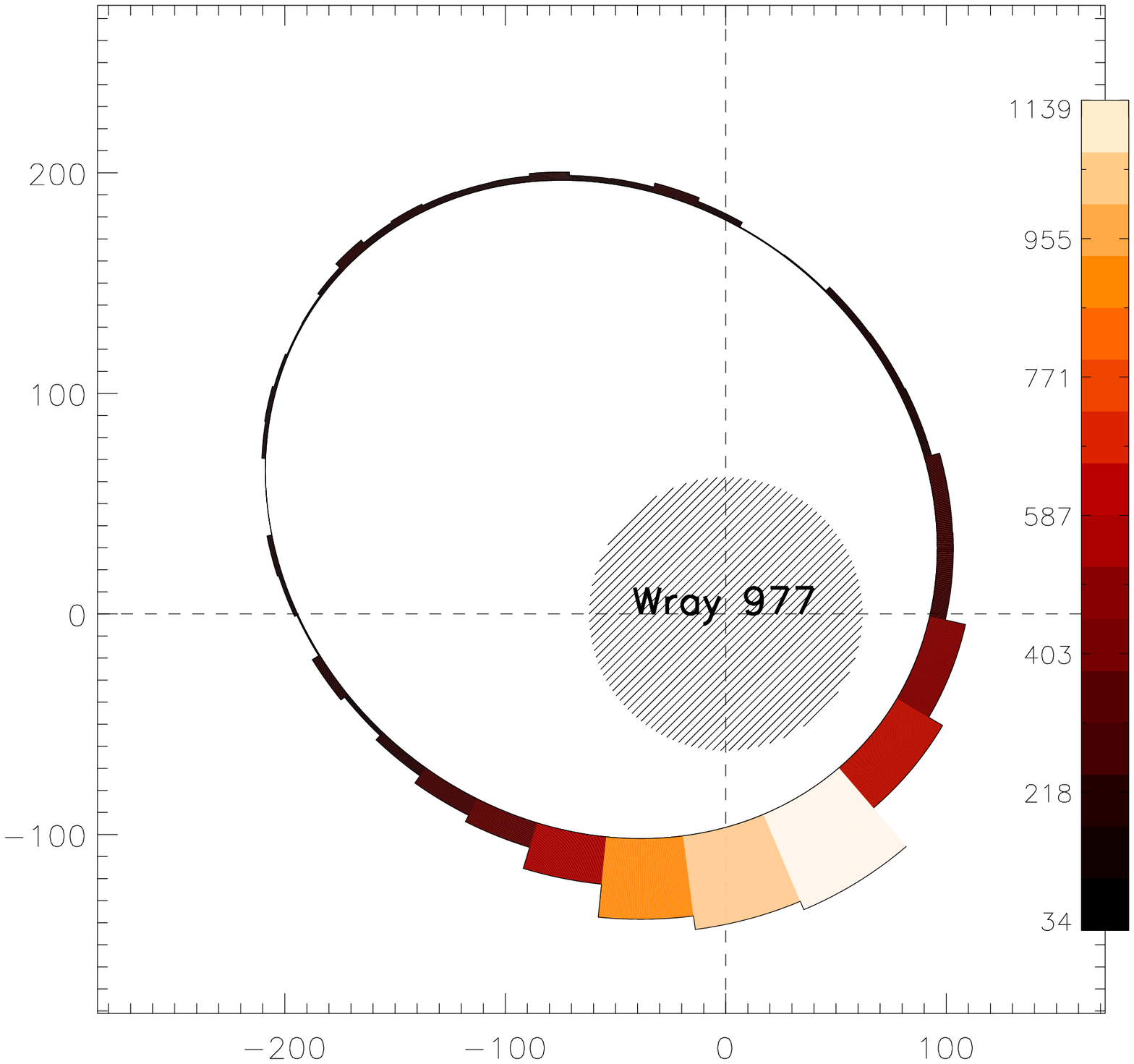}
 }
 \subfigure[]{
     \includegraphics[height=0.3\textheight]{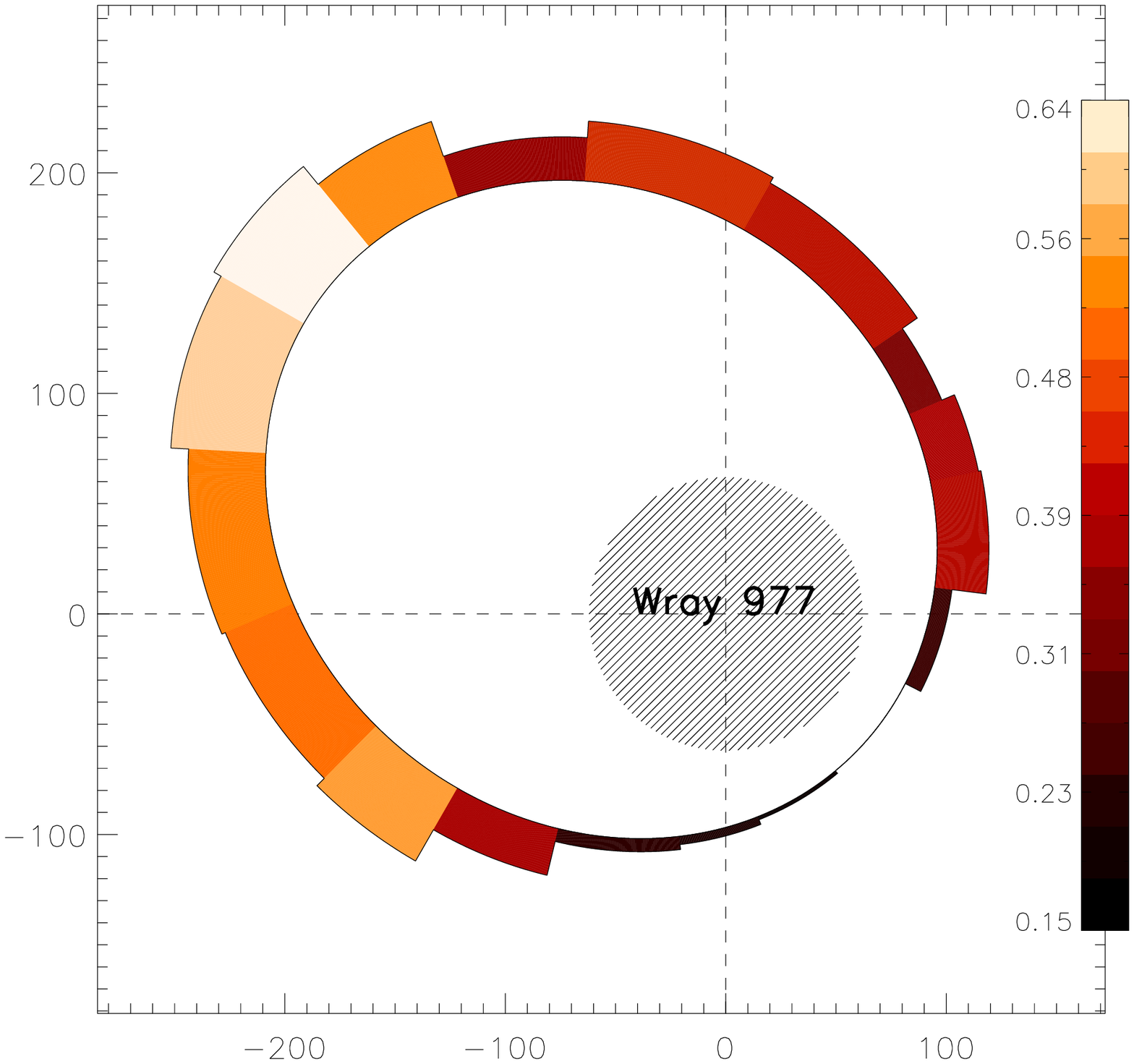}
 }
  \subfigure[]{
     \includegraphics[height=0.3\textheight]{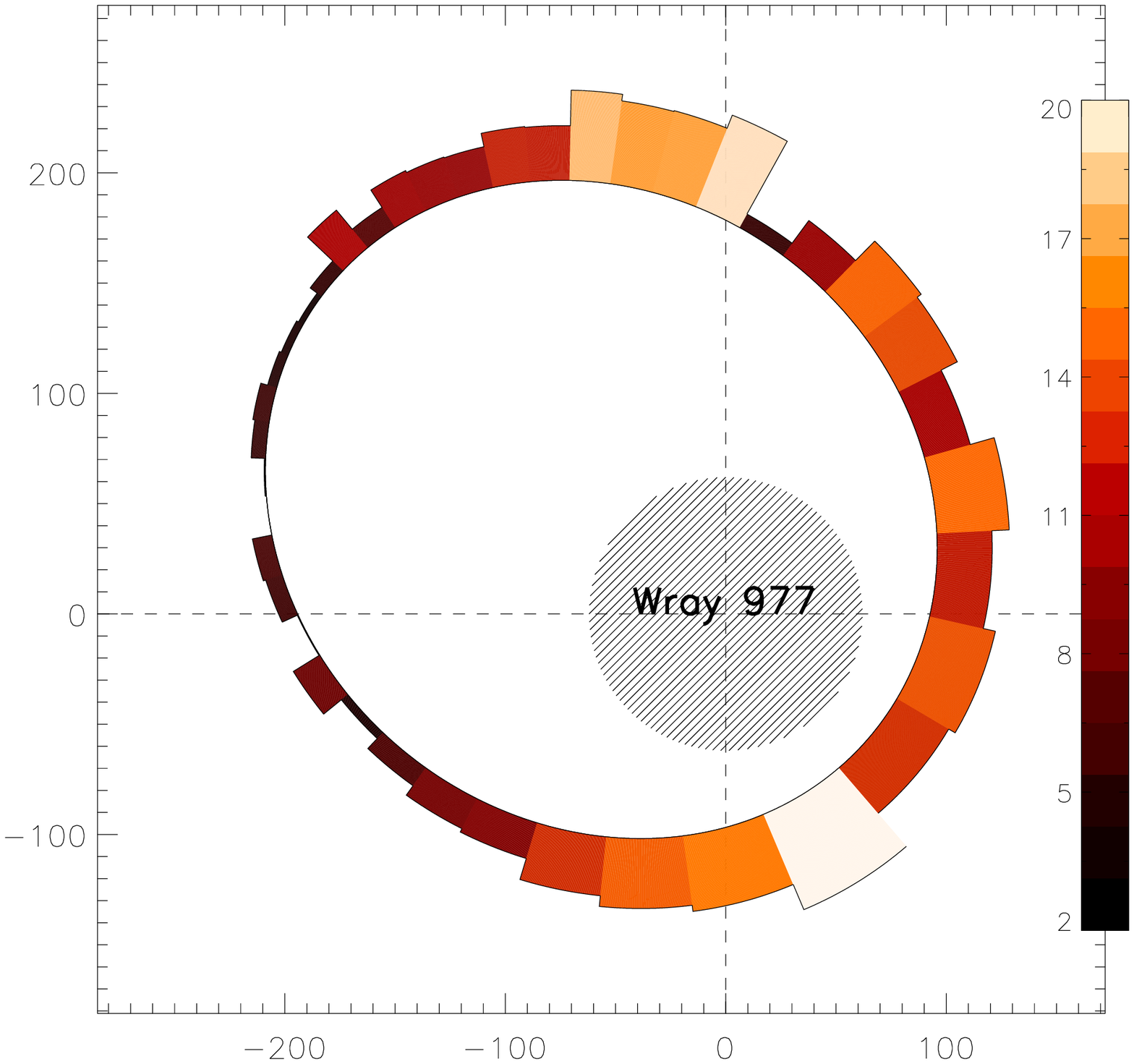}
 }

   \caption{  \label{fig:orbital_res}Summary of the orbital behavior
     of GX~301-2: hard X-ray flux (a), pulsed fraction (b) and
     hardness ratio (c). Units along the axis are in solar radii.}
\end{figure*}

The general behavior of the source is summarized in the
synoptic panels of Figure~\ref{fig:orbital_res}, where the parameters
are graphically displayed along the binary orbit. These plots show
that even if the only significant feature in the flux modulation
observed by SuperAGILE occurs at the pre-periastron phase, the
emission mechanism undergoes important changes also in other parts of
the orbit. The passage at the periastron is accompanied with a
hardening of the radiation and with a sudden and large decrease of
the pulsed fraction, as already reported by \cite{Labarbera2005},
using the {\it BeppoSAX} observations, and \cite{Lutovinov2009}
using {\it INTEGRAL} data. The pulsed fraction shows
(Figure~\ref{fig:pulsed_frac}(a)) a clear anti-correlation with the
source flux, thus presumably with the accretion rate. 
Instead, the hardness ratio does not show a clear dependence on the
X-ray flux (and then the accretion rate), suggesting a more complex
relation.
It is interesting to note that the largest
drop in the pulsed fraction is approximately a factor of 3, from
$\sim$0.55 to $\sim$0.2, while the flux increase in the same phase
interval is about a factor of 10. This implies that the excess hard
radiation emitted during the pre-periastron flare does not represent a
proportional luminosity increase of the same regions emitting in
the preceding orbital phases, that would leave the pulsed fraction
unchanged, nor an unrelated emission from a site not subject to
occultation by the NS upon spinning, that would cause a larger
drop in the pulsed fraction. Thus, either the site of the excess
emission is only partially occulted to the line of sight as the NS
rotates, or it is associated with an intervening radiative transfer
mechanism able to smear the coherence of the photon arrival times
(as proposed by \citealt{Makino1985}). Moving from the
periastron to the apastron, the pulsed fraction seems to recover
gradually to its maximum value. Instead, the hardness ratio shows a
further, smaller peak at phases near 0.25 mostly due to a depression
in the soft X-ray flux. Thus the cause intervening in obscuring or
depressing the emission of soft X-rays where the NS approaches the
apastron phase, leaves the persistent hard X-ray component unaffected,
as witnessed by the unchanging flux level measured by SuperAGILE.

\cite{Leahy2002} \citep[see also][]{Leahy2008} quite convincingly
modeled the orbital modulation of the soft X-ray emission in the
GX~301-2 system as driven by wind accretion plus a mass stream,
drawing an Archimedes spiral path and periodically intercepting
the neutron star. The effect of the spiral stream accretion on the
orbital light curve is the pre-periastron flare and the broad bump
around the apastron phase, both caused by the sudden and large
increase in the accretion rate. The larger density and narrower
geometrical size of the intercepted accretion stream near the
periastron phase causes a flare shorter and more intense than at
the apastron phase. In this scenario, the same matter that
accretes is also responsible for a very large absorption column
(thus the hardening) and for scattering of the radiation from the
neutron star, bringing loss of timing coherence (decrease in the
pulsed fraction). This is indeed what the ASM data show at soft
X-rays. 
In the same model, the wind component determines a smooth variation of
the flux, peaking only at the pre-periastron phase. This is what the
SuperAGILE data show at hard X-rays.

\cite{Leahy2008} show that around phase 0.25 the soft X-rays
are largely absorbed by a partial occultation of the neutron star
by the atmosphere of the companion star. This is consistent with
both the ASM flux modulation and with the SuperAGILE/ASM hardness
ratio. However, we observe an even more intense hardening peak of
the emitted radiation at the pre-periastron that is not shown in
the model prediction by \cite{Leahy2008}. In this case the
hardness peak is not due to a decrease in the soft X-ray flux, but
to a larger increase in the hard X-rays than in soft X-rays, still
in the presence of a strong flare in both energy bands. If
absorption is responsible for such an effect, this implies that
the site of the large flare lies behind the absorbing material
(thus the unabsorbed soft X-ray flux would be far larger) and both
the flare and absorption occur at the same orbital phase,
suggesting that the two are related in origin. Large increases
(factor of $\sim$10) in the absorption column near the periastron
phase were indeed reported by early studies (\citealt{Haberl1991} using
{\it EXOSAT} data, \citealt{Leahy1991} using {\it TENMA} data) while more recent
studies, using a partial covering absorber model, reported smaller
variations (factor of $\sim$2 with BeppoSAX, \citealt{Labarbera2005}) and
rather scattered in orbital phase (\citealt{Mukherjee2004}, with
{\it RXTE}/PCA). 

As already mentioned, at this phase we detect a large decrease in
the pulsed fraction, implying that the hard X-rays here are more
loosely affected by the NS spinning than the soft X-rays. This may
suggest that (at least) the hard X-ray emission at the
pre-periastron peak may be associated with temporary conditions
setting up when the NS crosses the accretion stream at its minimum
distance from the companion star. It is also interesting to remind
that just at this orbital phase, a detection of non-thermal radio
emission from GX~301-2 has been very recently reported by
\cite{Pestalozzi2009}.

Changes of the properties or sites of the hard X-ray
emission along the orbit are suggested also by our study of the
neutron star pulse profile as a function of the orbital phase. 
Similar to what found by
\cite{Labarbera2005} using the {\it BeppoSAX} data, we detected a large
variability in the pulse shape of the pulsar in the hard X-ray
range. The main points of variability we detected are: the structure
of the main peak, the intensity at the interpulse and the amplitude of
the pulsation (i.e., the pulsed fraction discussed above). As found by
{\it BeppoSAX}, in the SuperAGILE data the pulse appears smoothest at
orbital phases near the pre-periastron. The shape we obtained for the
orbital phase interval 0.91--0.96 is reminiscent of the {\it BeppoSAX}/PDS
curve. 
However, in approaching and leaving the phase of the large flare,
the pulse shape shows transient features. A small peak structure
(at spin phase around 0.1 in Figure~\ref{fig:pulse_shape_profile})
gradually appears on the rising and decaying phases of the flare
(orbital phases from 0.91 to 0.06), being more prominent after the
periastron. 
Simultaneously, the flux level at the inter-pulse phase (around
0.25) becomes higher. As the neutron star approaches the apastron
and goes beyond, the pulse shape at energies above 20~keV becomes
increasingly complex (although the smaller statistics does not
allow us to identify individual substructures), with the main
pulses becoming alternately narrower and wider. Interestingly,
approaching again the pre-periastron phase, a small peak appears
in the interpulse region. This clearly shows up at orbital phase
0.88--0.91. All of these features, including the smaller peak, are not
in contrast with the {\it BeppoSAX} results, obtained by
\cite{Labarbera2005} in their detailed analysis of the pulse shape in
four short orbital phase intervals. Altogether these results seem to
suggest that the emitting region on the neutron star is
significantly affected by the changing accretion pattern along the
binary orbit.

One last consideration derives from the secular variation of the
pulse period of the neutron star, collected in
Figure~\ref{fig:spin_history}. As already commented and discussed by
several authors \citep[e.g.][]{Koh1997}, the spin history is complex
and shows several changes in the spin-up and spin-down trends,
sometimes also very rapid. The major changes in the past were probably
accompanied to important variations in the emission. In the last
$\sim$15 years (from MJD~49500 onward), a general spin-down trend is
settled,  perturbed only by few short and rapid spin-up episodes. In
particular, despite some small-amplitude timing noise that SuperAGILE
detected (Figure~\ref{fig:spin_history}(b)), in the last $\sim$two years
the GX~301-2 system appears rather steadily evolving with a spin-down
rate of about 1.3~s~yr$^{-1}$, showing no rapid or large
perturbations. 

\section{Summary and conclusions}
\label{sec:conclusion}

We studied the long term behavior of the HMXB GX~301-2 using the
hard X-ray data collected by SuperAGILE during the first year of
the {\it AGILE} mission. The observations, carried out in the 20--60~keV
energy band, have a large net exposure, about 3.7~Ms, and cover
all the orbital phases of the binary system. The SuperAGILE data
offer one of the most complete hard X-ray monitoring of the
41.5-day long binary period available to date. The data span
covers large fractions of six orbital cycles. The secular trend of
the $\sim$680 s pulse period is consistent with the previous
observations, although the SuperAGILE data show a decrease in the
spin-down rate with respect to the observations reported by
\cite{Doroshenko2008}. The spin-down trend is approximately
constant, but there is indication of a significant timing noise, of
the order of $\sim$0.2\%.

Complementing the SuperAGILE data with those of {\it RossiXTE}/ASM, the
source behavior was characterized at all the phases of the binary
orbit, in terms of its soft and hard X-ray flux, spectral
hardness, spin period history, pulsed fraction and pulse shape.
Compared with the soft X-ray data, the hard X-ray lightcurve
exhibits a smaller modulation of the emission during the orbital
phases 0.1--0.8 and a larger intensity of the pre-periastron flare.
This results in a spectral hardness which presents clear orbital
modulation, with peaks in correspondence with the pre-periastron
flare and near phase 0.25. The timing analysis of the hard X-ray
emission showed a variable pulse shape profile as a function of
the orbital phase, with substructures detected near the passage at
the periastron, and a clear modulation of the pulsed fraction,
which appears in turn strongly anti-correlated with the source
intensity.

Overall, the scenario depicted by the SA data appears generally
consistent with the model for this binary system proposed by
\cite{Leahy2008}. The predictions of the model are based on the
variability in the accretion rate caused by the orbital motion of
the NS in the environment of the matter outflow from the companion
star. The model is suited to the low energy observations
by {\it RXTE}/ASM, for which it reproduces the peaked flare at the
pre-periastron and the broader peak at the apastron. They both are
related to the NS crossing of the accretion stream, the difference
between the two being mostly in the local density and geometrical
size of the stream at the crossing point. The higher accretion
rate causes a soft X-ray brightening, and scattering of the
radiation by the same matter would cause loss of coherence and
then a decrease in the pulsed fraction. In the hard X-rays we do
not detect the broad peak at the apastron phase (although other
hard X-ray observations with BATSE and {\it Swift}/BAT suggest time
variability for it). Also, our observation of the folded hardness
ratio (20--50~keV to 2--12~keV) along the orbit is not fully
consistent with the model expectations. Indeed the model predicts
an increase in the column density near the apastron phase, but it
does not predict a similar increase near the pre-periastron, that
we instead clearly detect.

More in general, our results fit in the overall scenario of the
binary X-ray pulsars \citep[e.g.][]{Parmar1989}, where
the timing properties are related to the source luminosity. As a
matter of fact, our analysis shows that the pulsed fraction is clearly
anti-correlated with the source intensity, allowing us to attribute
the accretion rate the role of driving parameter in the system.
However, while the timing properties appear rather well correlated
with the source intensity, the same is not true for the hardness
ratio, for which we cannot identify a trend with the flux. This
implies that the emitting process and/or geometry change along the
orbit in a way that is not immediately associated to the variation
of the accretion rate only. 

\acknowledgements
We thank L. Stella, P. Casella and R. Campana for useful
discussions and for support in data interpretation.
The {\it AGILE} Mission is funded by the Italian Space Agency (ASI) with scientific
and programmatic participation by the Italian Institute of Astrophysics (INAF)
and the Italian Institute of Nuclear Physics (INFN). We acknowledge the use of
public data from the {\it RXTE}/ASM data archive.

{\it Facilities:} \facility{AGILE}, \facility{RXTE (ASM)}

\end{document}